\documentclass[a4paper,11pt]{article}
\usepackage{jheppub} 
\usepackage{lineno}
\usepackage{appendix}
\usepackage{amsthm}
\usepackage{lineno}
\usepackage{bm}
\usepackage{braket}
\usepackage{amsmath}
\usepackage{amsfonts}
\usepackage{mathtools}
\usepackage{graphicx}
\usepackage{cancel}
\usepackage{slashed}
\usepackage{wrapfig}
\usepackage{nextpage}
\usepackage{enumitem}
\usepackage{float}
\usepackage{hyperref}
\usepackage{amssymb}
\usepackage{enumitem}
\hypersetup{colorlinks=true, urlcolor=black, citecolor=blue, linkcolor=black}

\usepackage{graphicx} 

\newcommand{\Tr}{\mathrm{Tr}}

\title{\boldmath Fuzzy Black Holes from Mass Generation in Matrix Compactification}

\author[1]{Davide Laurenzano\note{Corresponding author.}}
\author{and John F. Wheater}

\affiliation[a]{Rudolf Peierls Centre for Theoretical Physics, University of Oxford \\ Parks Road, Oxford OX1 3PU, United Kingdom}

\emailAdd{davide.laurenzano@physics.ox.ac.uk, john.wheater@physics.ox.ac.uk}

\abstract{
We investigate a mechanism for generating mass terms in the IKKT and BFSS matrix theories through compactification on a torus and the derivation of a zero-mode effective theory, emphasising the crucial role of fermionic boundary conditions. Extending a recent proposal developed for the IKKT model to the BFSS framework, we explore a broader class of mixed fermionic boundary conditions in both theories. This choice leads to a distinct effective theory with intermediate features, where a mass term is generated together with fermionic zero modes. In the BFSS case, this setup further allows for the construction of black hole solutions. The resulting geometry takes the form of a fuzzy sphere, with quantum excitations in the fermionic sector accounting for the corresponding black hole entropy.
}

\begin{document}
\maketitle

\section{Introduction}
The search for a quantum framework which unifies all fundamental interactions, including gravity, has long been a central topic of research in Theoretical High Energy Physics. Indeed, the most promising candidate for such a framework is string theory. Perturbative formulations of string theory have, in recent years, led to many theoretical breakthroughs. One of the main difficulties that ongoing efforts to derive testable predictions from string theory have to address is the fact that the space-time string theory lives in is constrained to be ten-dimensional. Hence, constructing observables proper to the four-dimensional space-time we live in requires dealing with compactification procedures. 

Among the shortcomings of string theory is that it only provides a perturbative expansion in the string coupling $g_s$, and the theory is not well defined at strong coupling. Moreover, it assumes a fixed background, whereas we would like a quantum theory of gravity to be background independent; in fact, spacetime itself should be dynamical. The presence of dualities connecting all five consistent superstring theories, as well as eleven-dimensional M-theory (the unknown theory which arises as the strong coupling limit of type IIA), hints at the existence of a single, more fundamental theory which encompasses all. With these motivations, non-perturbative formulations of string/M theory have been proposed, which take the form of matrix theories \cite{Ishibashi:1996, Banks_1996, Susskind:1997, Berenstein:2002}. These theories provide a framework where space or spacetime emerges from fundamental matrix degrees of freedom. 

Compelling evidence for the emergence of $1+3$ macroscopic dimensions out of the $1+9$-dimensional spacetime that string theory naturally inhabits would constitute significant progress in non-perturbative string theory. Progress in this direction has been made for the IKKT model in which it has been shown numerically and analytically that a time-direction can be defined such that spontaneous symmetry breaking of Lorentz symmetry from \(SO(10)\) to $SO(4) \times SO(6)$ occurs in the Euclidean model \cite{Nishimura:2011, Anagnostopoulos:2020}. Moreover, numerical studies of the Lorentzian theory have hinted at a similar mechanism \cite{Nishimura:2012}; the symmetry breaking pattern is $SO(1,9) \rightarrow SO(1,3) \times SO(6)$ and three space directions become macroscopic after a critical time, whereas the remaining six are of Planck size. Work is in progress \cite{Anagnostopoulos:2022} to remove some problematic technical assumptions, which are needed to avoid the sign problem in lattice simulations of the model. Such evidence has motivated studies of matrix cosmology \cite{Brahma:2021, Brahma:2022, Brahma:2022hjv, Brahma:2022ikl}, in which it is proposed that early universe cosmology and spacetime metric emerge from the matrix model. 

A key requirement for a $1+3$-dimensional spacetime to emerge is that a mass term, which serves as a regulator, is introduced into the theory \cite{Nishimura:2012}. More generally, cosmological solutions were found in various analyses of IKKT with a mass term \cite{Kim:2012mw, Hatakeyama:2019jyw, Steinacker:2017bhb, Sperling:2018xrm, Karczmarek:2022ejn}. Naturally, the question arises as to the nature of this mass term. An interesting mechanism was proposed in \cite{Laliberte_2024}, where compactification of IKKT on a six-torus is considered. Upon choosing anti-periodic boundary conditions for fermions, it is shown that, for a large radius of compactification, supersymmetry is broken completely; the effective theory of zero winding modes is purely bosonic and contains a mass term. This explains the origin of the mass deformation.

Here, we first generalise this result by considering mixed boundary conditions for fermions, where some degrees of freedom are given periodic boundary conditions, and others are given anti-periodic ones. As we will see, we will be able to obtain an effective theory that both retains some fermionic degrees of freedom and generates a mass term. The presence of fermionic degrees of freedom is an important feature of the theory; indeed, Monte Carlo results show that they are crucial for symmetry breaking to occur \cite{Nishimura:2011}.

The problem of establishing a symmetry-breaking pattern which entails the emergence of spacetime in BFSS is made more complicated by the difficulties in simulating the model. However, some analytical results \cite{Brahma:2022ifx} suggest using a Gaussian expansion method \cite{Kabat_2000} that symmetry breaking does occur in BFSS. Despite this success, there is no conclusive evidence that the symmetry-breaking pattern is the one expected for the emergence of three large space directions, namely\footnote{One of the main differences between the two theories is that in BFSS, time is a variable and does not emerge. The space symmetry is then $SO(9)$.} $SO(9)\rightarrow SO(3)\times SO(6)$. Hence, by analogy with IKKT, we are motivated to generalise the result for mass generation to BFSS theory, as a further step towards the achievement of such a symmetry-breaking pattern in BFSS. Since compactifications of BFSS theory on a six-torus are problematic \cite{Seiberg:1997, Sen:1997}, we will instead consider compactifications on a complementary three-torus. We will show that the result is qualitatively analogous, and that anti-periodic boundary conditions for fermions still yield an effective theory for winding zero modes which presents a mass term and no fermions. 

We will consider a class of mixed boundary conditions and show how they result in a theory which contains both a mass term and $SO(3)$ Weyl fermions. Such a theory shares some features with a recent bottom-up proposal \cite{Chu2024, Chu:2024edh} of a matrix model whose solutions appear to be suitable to characterise black holes. One noticeable feature is that the fermionic sector accounts for the black hole entropy \cite{Chu:2023}. In common with that proposal, the theory we derive through compactification, namely a non-supersymmetric mass-deformed BFSS theory with half the fermionic degrees of freedom, also admits three-dimensional fuzzy sphere solutions as classical solutions for the background spacetime. 

Fuzzy sphere solutions of matrix theories have been considered in different settings. In the context of BMN theory, they have been proposed as the holographic dual to supergravity on the supersymmetric plane-wave background \cite{Berenstein:2002, Lin:2004kw}. Furthermore, in BFSS theory, they have been considered in relation to black hole physics \cite{Kabat:1997, Iizuka:2013, Hyakutake:2018}. Here, we construct black hole solutions starting from a fuzzy sphere background, along the lines of \cite{Chu2024}. One of the main results of the present paper is that the fermionic part of the theory, albeit having a different interaction term from \cite{Chu2024}, is still able to account for the black hole entropy. In particular, we will see that the entropy computed in this way scales like $S \sim N^\beta$, $\beta \leq 1$, $N$ being the matrix size, in agreement with previous studies of black holes in matrix theory \cite{Banks:1997, Banks:1997hz, Horowitz:1997}. The saturated value $S \sim N$ signals the Gregory–Laflamme instability \cite{Gregory:1993, Gregory:1994} that leads to the black hole/black string transition. Hence, this work constitutes an explicit top-down construction of a Schwarzschild black hole from compactifications of matrix theory, where the Bekenstein-Hawking entropy is obtained as the degeneracy of the quantum state of fermionic degrees of freedom.

This paper is organised as follows: in section \ref{sec:ReviewIKKT} we review the results of \cite{Laliberte_2024} about mass generation from compactification of IKKT theory. Section \ref{sec:IKKT_Mixed} generalises these results, introducing a new class of boundary conditions for fermions, leading to an effective theory with mixed features. We generalise the results for completely periodic/anti-periodic fermionic boundary conditions to BFSS in section \ref{sec:BFSSComp}, where we carry out the explicit computation of the zero-mode effective theory. We then consider mixed boundary conditions in BFSS in section \ref{sec:BFSS_Mixed}, which leads to an interesting effective theory where a mass term is present alongside fermionic degrees of freedom. We leverage such a theory in section \ref{sec:BH}, where we explicitly construct a black hole solution from fermionic excitations on top of a bosonic fuzzy sphere background. We will then comment about the mass-to-radius and the entropy-to-radius relations obtained in this way, and show how their consistency requires tuning the scaling of the initial torus radius with $N$. Lastly, we comment on our results in section \ref{sec:Conclusion}.  
\section{Review of IKKT compactification}
\label{sec:ReviewIKKT}
In this section, we review the results of \cite{Laliberte_2024} on the compactification of IKKT matrix theory on a six-torus $\mathbb{T}^6$. For ease of comparison, we will adhere closely to the formalism and notation of \cite{Laliberte_2024}.  
However, for clarity and completeness, we also give an explicit construction in Appendix \ref{sec:torusdecomposition}.

The IKKT matrix theory is the large $\mathcal{N}$ limit of the matrix integral
\begin{equation}
    \mathcal{Z}= \int \mathcal{D}[A^I] \mathcal{D}[\psi_{\alpha}] \,e^{iS_{IKKT}} \, ,
    \label{eq:MatInt}
\end{equation}
where $A^I$, $I=0,...,9$ and $\psi_{\alpha}$, $\alpha=1,...,32$ are $\mathcal{N}\times \mathcal{N}$ matrices, the latter being Grassmann valued. Here $S_{IKKT}$ is the action given by
\begin{equation*}
    S_{IKKT} = -\frac{1}{2g^2}\Tr\left[\frac{1}{2}[A^I, A^J]^2+\bar{\psi}\Gamma^I[A_I, \psi]\right] \, ,
\end{equation*}
which enjoys a global $SO(1,9)$ symmetry and $\Gamma^I$ are the corresponding gamma-matrices in 1+9-dimensions. The bosonic matrices $A^I$ form a vector of $SO(1,9)$. In the physical interpretation of the theory, these encode information about space-time coordinates in 1+9 dimensions. In particular, diagonal entries encode the location of \(N\) \(D_{-1}\) instantons, whereas off-diagonal degrees of freedom describe open strings stretched between two branes. The fermionic matrices $\psi$ are spinors of the same group, which act as super-partners of the bosons, such that the theory is maximally supersymmetric. In Lorentzian signature, inner products are computed with respect to the metric $g_{IJ}=diag(1, -1,...,-1) $, $I, J=0,...,9$. In the following, it will be useful to Wick rotate to Euclidean signature through the replacement $A^0 \rightarrow iA^0$, $\Gamma^j \rightarrow i\Gamma^j$, $j=1,...,9$, so that the Euclidean action reads 
\begin{equation}
    S^E_{IKKT} = -\frac{1}{2g^2}\Tr\left[\frac{1}{2}[A^I, A^J]^2+i\bar{\psi}\Gamma^I[A_I, \psi]\right] \, .
    \label{eq:EuclideanIKKT}
\end{equation}
Note that, as pointed out in \cite{Hartnoll:2024}, in Euclidean signature, one cannot impose a reality condition on spinors consistently with supersymmetry. In this case, the notation $\bar{\psi}_{\beta}:= (\psi^T C_{10})_{\beta}=\psi^\alpha C_{10, \alpha \beta}$, where $C_{10}$ is the charge conjugation matrix in ten dimensions, should be regarded as a convenient definition. In fact, while in Lorentzian signature the fermionic degrees of freedom consist of one 32-component Weyl $SO(1,9)$ spinor satisfying a Majorana reality condition, in the Euclidean theory only the $SO(10)$ Weyl spinor $\psi$ appears in the action and not its complex conjugate, and the path integration is restricted to Hermitian matrices. Of course, the overall number of degrees of freedom is the same in the two cases. 

We consider the case where the target space of the theory is compactified on a six-dimensional torus, where all directions have the same extent $2\pi L$. The construction is inspired by a technique which was first developed in the context of compactifications of the BFSS matrix model \cite{Banks_1999}. The same mechanism was later deployed in the context of IKKT in order to compactify the theory on a thermal circle \cite{Laliberte_2023}. The main idea is to consider the existence of unitary operators $U^a$, $a=4,...,9$, which generate translations in the six target space directions we wish to compactify. If one further imposes that such operators commute with each other when they act on different directions, then translations can be generated independently. 

The action of the $U^a$ operators on fermionic matrices appearing in (\ref{eq:EuclideanIKKT}) can be either periodic or anti-periodic when circling around any of the torus directions. For the moment, we consider the fully anti-periodic case, for which the action of translation operators reads
\begin{equation}
\begin{aligned}
    (U^a)^{-1}A^{\mu} U^a&=A^{\mu} \\
    (U^a)^{-1}A^{b} U^a&=A^{b}+2\pi L \delta^{ab}\\
    (U^a)^{-1}\psi U^a&=-\psi \, \, .
\end{aligned} \qquad 
\begin{aligned}
    \mu&=0,...,3\\
    a&=4,...,9
\end{aligned}
\label{eq:Torustranslation}
\end{equation}
 Since matrices $A^I$ encode information about target space locations, the transformation (\ref{eq:Torustranslation}) can be thought of as wrapping around the torus once. We can think of every compact target space direction as a real line where consecutive intervals of length $2 \pi L$ are identified; then the transformation (\ref{eq:Torustranslation}) translates one interval into the next. The Lagrangian of the theory obtained in this way will be invariant under the action of \(U^a\). Note that this is true even for anti-periodic fermions, since fermionic fields appear in the Lagrangian in pairs.
 
 This construction explicitly implements the method of mirror images introduced by Washington Taylor in \cite{Taylor:1996ik} (and recently used, for example, to compute the three graviton amplitude \cite{Herderschee:2023pza}, which enabled the proof of a Soft Theorem in BFSS \cite{Herderschee:2023bnc}\footnote{See \cite{Laurenzano:2025ywy} for an alternative approach to Soft Theorems in BFSS theory.}).
The relations (\ref{eq:Torustranslation}) are realised by viewing the Hilbert space of the matrices as a tensor product 
\begin{equation}
    X=Y \otimes Z \, ,
    \label{eq:TensorHilbert}
\end{equation}
where $Y$ is the space of $N \times N$ matrices which is left invariant by the translation and $Z$ is the space of $M \times M$ matrices acted upon by the translation operator (hence the original matrix size $\mathcal{N}=N \times M$). Introduce a pair of $M \times M$ matrices $p^a$ and $q^a$, satisfying $[q^a, p^b]=i \delta^{ab}$. (Note that this only makes sense in the limit of $M \rightarrow \infty$, and we implicitly assume that we work in this limit.)
Then, to implement the relations \eqref{eq:Torustranslation}, the operators $U^a$ are given by 
\begin{equation}
    U^a= \mathbb{I}_N \otimes e^{-2\pi i q^a} e^{-i p^a} \, ,
\end{equation}
and the matrix degrees of freedom by
\begin{equation}
    \begin{aligned}
        A^{\mu}&=\sum_{n^b \in \mathbb{Z}^6}A^{\mu} (n^b) \otimes e^{in^b p^b} \\
        A^{a}&=\sum_{n^b \in \mathbb{Z}^6}A^{a} (n^b) \otimes e^{in^b p^b} +2\pi L \, \mathbb{I}_N \otimes q^a \\
        \psi&=\sum_{r^b \in \mathbb{L}_6 }\psi (r^b) \otimes e^{ir^b p^b} \, .
    \end{aligned}
    \label{eq:Toepldecomposition}
\end{equation}
Here $\mathbb Z^6$ is a six-dimensional hypercubic lattice with vertex positions labelled by integers, and
\begin{equation}
    \label{eq:shitlattice}
\mathbb{L}_6 = \left(\mathbb{Z}+\frac{1}{2}\right) \times ... \times \left(\mathbb{Z}+\frac{1}{2}\right) 
\end{equation}
is a six-dimensional hypercubic lattice with vertex positions labelled by half-integers.

Essentially, periodic/anti-periodic boundary conditions are imposed by requiring that winding numbers $n^b, r^b$ take integer/half-integer values, respectively.
In Appendix \ref{sec:torusdecomposition}, we provide an explicit formulation of this prescription, where the operators appearing in the decomposition of matrices and in the translation operator are realised as matrices in the \(|q^a\rangle\) basis, both for periodic and anti-periodic fermions.

Physically, along each compact direction, \(D_{-1}\) instantons with coordinates \(r^a\) in a fundamental region \(r^a \in [0, 2\pi L]\) are replicated an infinite number of times. Open strings connecting them can have winding number zero, when they stretch across two branes in the same fundamental region, or non-zero, in which case they connect two copies living in different regions. The matrices (\ref{eq:Toepldecomposition}) can be represented in block form, where diagonal blocks describe fundamental regions, and off-diagonal blocks account for winding modes. See \cite{Laliberte_2024} for more details. 

The consequence of the compactification procedure is a theory whose degrees of freedom wind around the torus directions. We name the resulting action the compact IKKT action. This is obtained by first defining the Fourier decomposition of fields in winding modes in terms of a six-dimensional momentum variable $\sigma$
\begin{equation}
    A^I(\sigma)= \sum_{n^b \in \mathbb{Z}^6} A^I(n^b) e^{2\pi i Ln^b \sigma^b}, \quad \psi(\sigma)= \sum_{r^b \in \mathbb{L}_6} \psi(r^b)e^{2\pi i Lr^b \sigma^b}, \quad \sigma^b \in [0, L^{-1}] \, .
    \label{eq:decomposition}
\end{equation}
Inserting the decomposition (\ref{eq:Toepldecomposition}) into the original IKKT action, tracing over the $Z$ subspace, and making use of the property
\begin{equation}
    \Tr[e^{i(n \pm m)^b p^b}]=M \delta_{n,\mp m}\, ,
    \label{eq:traceproperty}
\end{equation}
we find
\begin{equation}
    S_C= \frac{1}{2 g^2_{eff}} \int \frac{d^6 \sigma}{(2\pi L)^{-6}}\Tr \left[\frac{1}{2}F_{ab}F^{ab}+D_aA_\mu D^a A^\mu-\frac{1}{2}[A^\mu, A^\nu]^2+\bar{\psi}\Gamma^aD_a\psi-i \bar{\psi}\Gamma^\mu [A_\mu, \psi]\right]\, ,
    \label{eq:CompactAction}
\end{equation}
where the effective coupling is $g^2_{eff}=\frac{g^2}{M}$. In the above expression, we have dropped the dependence of matrix fields on \(\sigma^b\) for simplicity of the notation, and we recall that all the fields are \(N \times N\) matrices acting on the subset \(Y\) of the Hilbert space (\ref{eq:TensorHilbert}). 
The goal of the compactification procedure is to provide a mechanism for mass generation. To do so, we need to consider the decompactification limit and, in particular, the effective theory of zero-modes in this limit. The appearance of mass terms could lead to the breaking of the global $SO(10)$ symmetry. It will be useful to 
consider a decomposition of spinors which reflects the symmetry-breaking pattern we seek. Before turning our attention to the derivation of the effective theory, we review the spinor decomposition adopted in \cite{Laliberte_2024}, as this will be central to our discussion of boundary conditions in the next section. 

The decomposition of gamma matrices that we will adopt highlights a separation between $SO(4)$ (recall that we Wick rotated to Euclidean signature) and $SO(6)$ degrees of freedom. We consider the gamma matrices
\begin{equation*}
    \Gamma^\mu =\Tilde{\Gamma}_7\otimes \gamma^\mu, \quad \Gamma^a=\Tilde{\Gamma}^a \otimes \mathbb{I} \, ,
\end{equation*}
where $\Tilde{\Gamma}^a$ are $8 \times 8$ gamma matrices of $SO(6)$ and $\Tilde{\Gamma}_7$ is the corresponding chirality operator. The $SO(4)$ gamma matrices are further decomposed in the Weyl basis 
\begin{equation*}
  \gamma^\mu =  \begin{pmatrix}
        0 & \sigma^\mu \\
        \Bar{\sigma}^\mu & 0
    \end{pmatrix}\, ,
\end{equation*}
where $\sigma^\mu=(\mathbb{I}, \sigma^i)$, $\Bar{\sigma}^\mu=(\mathbb{I}, -\sigma^i)$ and $\sigma^i$ are Pauli matrices. The 10-dimensional chiral operator and charge conjugation decompose as 
\begin{equation}
\label{eq:GCC}
    \Gamma_{11}=\tilde{\Gamma}_7 \otimes \begin{pmatrix}
        \mathbb{I}_2 & 0 \\
        0 &-\mathbb{I}_2
    \end{pmatrix}, \quad C_{10}=C_6 \otimes \begin{pmatrix}
        i \sigma_2 & 0 \\
        0 & -i \sigma_2
    \end{pmatrix} \, ,
\end{equation}
where $C_6$ is the charge conjugation matrix in 6 dimensions.  It follows that the 32-component spinor $\psi$ can be written as 
\begin{equation}
    \psi=\begin{pmatrix}
        \psi^A_+ \\
        \psi^A_-
    \end{pmatrix}, \qquad A=1,2 \, ,
    \label{eq:SpinorIKKT}
\end{equation}
where $\psi^A_{\pm}$ are the two chiral components and each component is an 8-dimensional spinor acted upon by the $SO(6)$ Clifford algebra. 

The spinor decomposition can be plugged into the compact action (\ref{eq:CompactAction}), yielding 
\begin{multline}
    S_C= \frac{1}{2 g^2_{eff}} \int \frac{d^6 \sigma}{(2\pi L)^{-6}}\Tr \left[\frac{1}{2}F_{ab}F^{ab}+D_aA_\mu D^a A^\mu-\frac{1}{2}[A^\mu, A^\nu]^2+\bar{\psi}^A_+\tilde{\Gamma}^a\partial_a\psi^A_+ +\bar{\psi}^A_-\tilde{\Gamma}^a\partial_a\psi^A_- \right.\\
    \left.-i\bar{\psi}^A_+\tilde{\Gamma}^a [A_a, \psi^A_+]-i\bar{\psi}^A_-\tilde{\Gamma}^a [A_a, \psi^A_-]+i\bar{\psi}^A_+(\sigma^\mu)^{AB}[A_\mu, \psi^B_-]-i\bar{\psi}^A_-(\bar{\sigma}^\mu)^{AB}[A_\mu, \psi^B_+]\right] \, ,
    \label{eq:Compact}
\end{multline}
along with the ghost theory associated with the Lorentz gauge condition $\partial_a A^a=0$:
\begin{equation}
    S_{gh}=\frac{1}{g^2_{eff}} \int\frac{d^6 \sigma}{(2\pi L)^{-6}}\Tr \left[\partial^a \bar{c}D_a c \right] \, ,
    \label{eq:ghost}
\end{equation}
which arises from a gauge fixing delta function in the measure of the path integral (\ref{eq:MatInt}).

The idea is to look at the effective theory in the limit $L \gg g^{1/2}_{eff}$. By inserting the winding mode decomposition (\ref{eq:decomposition}) into the compact action (\ref{eq:Compact}) and ghost theory (\ref{eq:ghost}), we find that non-zero winding modes on the torus acquire a mass proportional to the torus size $2 \pi L$. This explicit mass term takes the form 
	\begin{multline*}
	S_{m}=\frac{1}{2g^2_{eff}}\sum_{{n^a \in \mathbb{Z}^6}}(2 \pi Ln^a)^2\Tr[A_I(-n^a)A^I(n^a)]\\+ \frac{1}{2g^2_{eff}} \sum_{r^a \in \mathbb{L}_6} (2\pi Li r^a )\Tr[\bar{\psi}_{+}^{A}(r^a)\tilde{\Gamma}^a \psi_{+}^{A}(r^a)]+ \frac{1}{2g^2_{eff}} \sum_{r^a \in \mathbb{L}_6} (2\pi Li r^a )\Tr[\bar{\psi}_{-}^{A}(r^a)\tilde{\Gamma}^a \psi_{-}^{A}(r^a)]\,.
	\end{multline*}
	 Note that the masses are proportional to the winding number. This implies, in particular, that the mass of open strings stretched between two well-separated branes, with coordinates \(x^J\) and \(y^J\) in the compact direction \(J\) and equal coordinates in the other directions, is \(m \propto|y^J-x^J+2\pi L n|\). Here \(n\) is the difference in winding number, and accounts for the stretching of the string when winding the compact dimension. It follows that, in the limit we are considering, such modes can be integrated out. The result of this operation is a Wilsonian effective theory for the zero modes. As explicitly shown in \cite{Laliberte_2024}, boundary conditions for fermions play a crucial role, and the effective theory obtained depends on the choice. Indeed, anti-periodic conditions for fermions break supersymmetry explicitly. However, this does not produce any divergences in the one-loop effective action, since the contributions arising from bosonic and fermionic non-zero modes running in the loop still cancel.

The outcome is an effective theory for zero-modes, which, after rotating back to Lorentzian signature, takes the form
\begin{equation*}
    S^0_{eff}=-\frac{1}{2g_{eff}^2}\Tr \left[\frac{1}{2}[A^I(0), A^J(0)]^2-M_{IJ}A^I(0)A^J(0)\right]\,,
\end{equation*}
where the mass matrix consists of two blocks
\begin{equation}
    M_{IJ}=\begin{pmatrix}
        \eta_{\mu \nu} M^2_4 & 0\\
        0 & \eta_{ab} M^2_6
    \end{pmatrix}\,,
    \label{eq:Massmatrix}
\end{equation}
and breaks the $SO(1,9)$ symmetry to $SO(1,3) \times SO(6)$. The two terms $M^2_4$ and $M^2_6$ in (\ref{eq:Massmatrix}) are given by
\begin{equation*}
    M^2_4=16(S_F-S_B)\frac{N}{L^2}, \qquad M^2_6=\frac{32}{3}(S_F-S_B)\frac{N}{L^2} \,,
\end{equation*}
where $S_F$ and $S_B$ are the infinite sums 
\begin{equation*}
    S_B=\sum_{n^b \in \mathbb{Z}^6}\nolimits' \frac{1}{(2\pi n^b)^2}, \qquad S_F=\sum_{r^b \in \mathbb{L}_6} \frac{1}{(2\pi r^b)^2}\,,
\end{equation*}
and the bosonic sum is performed over non-zero modes. These sums can be regularised, and the infinite part cancels in the difference, leaving a finite, positive number. Note that, because fermions do not have zero modes with this choice of boundary conditions, they completely decouple from the low-energy theory and supersymmetry is broken.

On the other hand, fully periodic boundary conditions for the fermions, in the form
\begin{equation*}
     (U^a)^{-1}\psi U^a=\psi \, ,
\end{equation*}
yield the low-energy action
\begin{equation*}
    S^0_{eff} = -\frac{1}{2g^2_{eff}}\Tr\left[\frac{1}{2}[A^I(0), A^J(0)]^2+\bar{\psi}(0)\Gamma^I[A_I(0), \psi(0)]\right] \, ,
\end{equation*}
which is identical to the non-compact IKKT action with an effective coupling $g_{eff}$. In particular, supersymmetry is completely intact, and no mass term is generated.

In the following sections, we will extend these computations to the case of IKKT with alternative boundary conditions and then to BFSS theory. 
\section{IKKT with interchanging boundary conditions}
\label{sec:IKKT_Mixed}
In this section, we consider the Wilsonian effective action obtained by integrating out non-zero modes from the compact IKKT action with mixed boundary conditions. We expect mixed boundary conditions to result in an effective theory which is an intermediate case between the fully supersymmetric/no mass case and the effective theory with a mass term and no fermions. Our goal is to obtain a theory which, as a consequence of compactification, features fermionic degrees of freedom, while still giving rise to a mass term. We consider two types of interchanging boundary conditions: the first swaps the two components with the same chirality, and the other exchanges the chiralities.
\subsection{Component reversing boundary conditions}
We begin by considering the spinor decomposition (\ref{eq:SpinorIKKT}). We impose that the Weyl spinor components are mapped into one another when wrapped around any direction on the torus, namely
\begin{equation}
    (U^a)^{{-1}}\psi^1_{\pm} U^a=\psi^2_{\pm}, \quad (U^a)^{{-1}}\psi^2_{\pm} U^a=\psi^1_{\pm} \, .
    \label{eq:reversedBC}
\end{equation}
This is obtained by rewriting the spinor in the form
\begin{equation*}
    \psi^A_{\pm}=\psi^1_{\pm}\begin{pmatrix}
        1 \\
        0
    \end{pmatrix}+\psi^2_{\pm}\begin{pmatrix}
        0 \\
        1
    \end{pmatrix}:=\frac{(\psi^{\uparrow}_{\pm}+\psi_{\pm}^{\downarrow})}{\sqrt{2}}\begin{pmatrix}
        1 \\
        0
    \end{pmatrix}+\frac{(\psi^{\uparrow}_{\pm}-\psi_{\pm}^{\downarrow})}{\sqrt{2}}\begin{pmatrix}
        0 \\
        1
    \end{pmatrix}=\frac{\psi^{\uparrow}_{\pm}}{\sqrt{2}}\begin{pmatrix}
        1 \\
        1
    \end{pmatrix}+\frac{\psi^{\downarrow}_{\pm}}{\sqrt{2}}\begin{pmatrix}
        1 \\
        -1
    \end{pmatrix}
\end{equation*}
and using the Toeplitz decomposition
\begin{equation}
    \begin{aligned}
        \psi^{\uparrow}_{\pm}&=\sum_{n^b \in \mathbb{Z}^6}\psi^{\uparrow}_{\pm}(n^b) \otimes e^{in^b p^b} \, ,\\
        \psi^{\downarrow}_{\pm}&=\sum_{r^b \in \mathbb{L}_6 }\psi^{\downarrow}_{\pm}(r^b) \otimes e^{ir^b p^b} \, \, ,
    \end{aligned}
    \label{eq:ToeplitzReversed}
\end{equation}
where we recall that $\psi_{\pm}^{\uparrow}$ and $\psi_{\pm}^{\downarrow}$ are 8-component spinors of $SO(6)$. We also recall that \(\mathbb{L}_6 \) is the lattice (\ref{eq:shitlattice}) of half-integer winding modes. The decomposition (\ref{eq:ToeplitzReversed}) highlights how every chiral component of the spinor has a part with periodic and a part with anti-periodic boundary conditions 
\begin{equation*}
    \psi^A_{\pm}=\Psi^{\uparrow}_{\pm}+\Psi^{\downarrow}_{\pm}, \qquad \begin{cases}
   \begin{aligned}
        \Psi^{\uparrow}_{\pm}&=\frac{\psi^{\uparrow}_{\pm}}{\sqrt{2}}\begin{pmatrix}
        1 \\
        1
    \end{pmatrix} \\
    \Psi_{\pm}^{\downarrow}&=\frac{\psi^{\downarrow}_{\pm}}{\sqrt{2}}\begin{pmatrix}
        1 \\
        -1
    \end{pmatrix}  
     \,.
    \end{aligned} 
    \end{cases}
\end{equation*}
Note that these boundary conditions are equivalent to the transformation generated by the operator
\begin{equation}
\label{eq:bcOp}
    T=\mathbb{I}_8 \otimes \begin{pmatrix}
        \sigma_1 & 0
        \\
        0 & \sigma_1
    \end{pmatrix} \, ,
\end{equation}
which acts on spinors when they are translated on a full circle around one of the torus directions. Following the spinor decomposition, the Weyl and Majorana \footnote{As pointed out in the previous section, only in the Lorentzian theory one can impose the reality condition, while in Euclidean signature the symmetry group is $SO(10)$ which does not admit a Majorana-Weyl representation.} relations 
\[
\Gamma_{11} \psi =\psi , \qquad \bar{\psi}=\psi^T C_{10}
\]
become
\[
\tilde{\Gamma}_7 \psi_{\pm}=\pm \psi_{\pm}, \qquad \psi_{\pm}=\pm i\sigma_2 C_6\bar{\psi}_{\pm}^T \, ,
\]
where we used (\ref{eq:GCC}). These are indeed preserved by the action of (\ref{eq:bcOp}), which transforms the chirality and charge conjugation matrices as 
\[
T\, \Gamma_{11}T^{-1}=\Gamma_{11}, \qquad T \, C_{10} T^{-1}=-C_{10} \, .
\]

The fermionic theory that we obtain, by inserting this decomposition with interchanging boundary conditions back into the IKKT action, reads 
\begin{multline*}
    S_F=\frac{1}{2 g^2_{eff}}\int \frac{d^6 \sigma}{(2\pi L)^{-6}} \Tr\left[\bar{\Psi}_{+}^{\uparrow}\tilde{\Gamma}^a\partial_a \Psi_{+}^{\uparrow}+\bar{\Psi}_{-}^{\uparrow}\tilde{\Gamma}^a\partial_a \Psi_{-}^{\uparrow}+\bar{\Psi}_{+}^{\downarrow}\tilde{\Gamma}^a\partial_a \Psi_{+}^{\downarrow}+\bar{\Psi}_{-}^{\downarrow}\tilde{\Gamma}^a\partial_a \Psi_{-}^{\downarrow} \right.\\
    \left.-i\bar{\Psi}_{+}^{\uparrow}\tilde{\Gamma}^a[A_a, \Psi_{+}^{\uparrow}]-i\bar{\Psi}_{-}^{\uparrow}\tilde{\Gamma}^a[A_a,\Psi_{-}^{\uparrow}]-i\bar{\Psi}_{+}^{\downarrow}\tilde{\Gamma}^a [A_a, \Psi_{+}^{\downarrow}]-i\bar{\Psi}_{-}^{\downarrow}\tilde{\Gamma}^a [A_a, \Psi_{-}^{\downarrow}] \right.
    \\
    \left.+i\bar{\Psi}^{\uparrow}_+ \sigma^\mu [A_\mu, \Psi^{\uparrow}_{-}]-i\bar{\Psi}^{\uparrow}_- \bar{\sigma}^\mu [A_\mu, \Psi^{\uparrow}_{+}]+i\bar{\Psi}^{\downarrow}_+ \sigma^\mu [A_\mu, \Psi^{\downarrow}_{-}]-i\bar{\Psi}^{\downarrow}_- \bar{\sigma}^\mu [A_\mu, \Psi^{\downarrow}_{+}]\right] \, .
\end{multline*}
Here, the fields have been expanded in winding modes as 
\begin{equation}
\begin{aligned}
     &A^I(\sigma)= \sum_{n^b \in \mathbb{Z}^6} A^I(n^b) e^{2\pi i Ln^b \sigma^b}\\
     &\Psi^\uparrow_{\pm}(\sigma)= \sum_{r^b \in \mathbb{Z}^6} \Psi^\uparrow_{\pm}(r^b)e^{2\pi i Lr^b \sigma^b}\\
     &\Psi^\downarrow_{\pm}(\sigma)= \sum_{r^b \in \mathbb{L}_6} \Psi^\downarrow_{\pm}(r^b)e^{2\pi i Lr^b \sigma^b} \\
     & c(\sigma)=\sum_{n^b \in \mathbb{Z}^6} c(n^b) e^{2\pi i Ln^b \sigma^b}\, ,
     \end{aligned}
     \label{eq:ModeExpansion}
\end{equation}
where we also indicated the mode expansion of ghost fields (with periodic boundary conditions) for later convenience. Note that the two spinors, with opposite boundary conditions, are completely decoupled from one another. This follows from the different nature of their winding modes, which causes terms of the form $\Bar{\Psi}^{\downarrow}_+ \sigma^\mu [A_\mu, \Psi^{\uparrow}_{-}]$ to vanish due to the property (\ref{eq:traceproperty}). The vanishing of these terms causes a loss of fermionic degrees of freedom running in diagram loops, ultimately because the transformation (\ref{eq:bcOp}) is not a symmetry of the compact action. We will show shortly that this results in generating a divergent mass term in the effective theory.

We proceed by rescaling the fields in the compact IKKT action, in order to make them dimensionless, using 
\begin{equation}
\label{eq:fieldsRescaling}
    A^I \rightarrow \lambda L A^I, \quad \psi \rightarrow \lambda L^{3/2} \psi, \quad c \rightarrow \lambda L c \,,
\end{equation}
where $\lambda$ is the dimensionless coupling constant $\lambda:= \frac{g_{eff}}{L^2}$, which will serve as the perturbative expansion parameter. Substituting the mode expansion (\ref{eq:ModeExpansion}) into the compact action yields 
\begin{equation*}
    S=S_0+S_{kin}+S_{int} \, ,
\end{equation*}
where $S_0$ is the zero-modes action, $S_{kin}$ is the action for kinetic terms of non-zero modes and $S_{int}$ are the interaction terms. In particular, the action for kinetic terms reads
\begin{multline}
    \label{eq:Kinetic}
    S_{kin}=\frac{1}{2}\sum_{{n^a \in \mathbb{Z}^6}}(2 \pi n^a)^2\Tr[A_I(-n^a)A^I(n^a)]+ \frac{1}{2} \sum_{r^a \in \mathbb{Z}^6} (2\pi i r^a )\Tr[\bar{\psi}_{+}^{\uparrow}(r^a)\tilde{\Gamma}^a \psi_{+}^{\uparrow}(r^a)]\\
    +\frac{1}{2} \sum_{r^a \in \mathbb{Z}^6} (2\pi i r^a )\Tr[\bar{\psi}_{-}^{\uparrow}(r^a)\tilde{\Gamma}^a \psi_{-}^{\uparrow}(r^a)]+ \frac{1}{2} \sum_{s^a \in \mathbb{L}_6} (2\pi i s^a )\Tr[\bar{\psi}_{+}^{\downarrow}(s^a)\tilde{\Gamma}^a \psi_{+}^{\downarrow}(s^a)]\\
    + \frac{1}{2} \sum_{s^a \in \mathbb{L}_6} (2\pi i s^a )\Tr[\bar{\psi}_{-}^{\downarrow}(s^a)\tilde{\Gamma}^a \psi_{-}^{\downarrow}(s^a)]+\sum_{{n^a \in \mathbb{Z}^6}} (2 \pi n^a)^2\Tr[\bar{c}(n^a)c(n^a)] \, ,
\end{multline}
which shows that non-zero modes are massive. We will use this fact to integrate them out at one-loop, in order to obtain the Wilsonian effective action for the zero-modes, 
\begin{multline}
    S^{eff}_0=-\ln\left(\prod_{n^a \in \mathbb{Z}^6}\nolimits'\prod_{r^b\in \mathbb{Z}^6}\nolimits'\prod_{s^c \in \mathbb{L}_6}\int [\mathcal{D}A(n^a)][\mathcal{D}\psi^{\uparrow}(r^b)][\mathcal{D}\psi^{\downarrow}(s^c)]e^{-S_C}\right) \\
    =S_0-\ln \left \langle e^{-S_{int}}\right \rangle-\ln Z_{kin} \, ,
    \label{eq:ZeroEff}
\end{multline}
where $\ln Z_{kin}$ is a constant, independent of the zero modes, which we discard. The second term is the correction to the zero-mode effective action. The expectation value is computed in perturbation theory (we will evaluate it at one-loop) with respect to the kinetic action (\ref{eq:Kinetic}). The interaction can be decomposed as 
\begin{equation}
    S_{int}=\sum_{i=1}^{11}V_i \, ,
    \label{eq:Sint}
\end{equation}
where $V_1, V_2$ and $V_{11}$ are purely bosonic or ghost terms, identical to the corresponding terms in the purely anti-periodic case \cite{Laliberte_2024}, whereas $V_i$ for $i=3,...,10$ include fermions and differ from  \cite{Laliberte_2024} due to the choice of boundary conditions. 

We first examine the bosonic terms, which are given by
\begin{equation*}
    \begin{aligned}
        V_1&=-\frac{\lambda^2}{4}\sum_{n^a, m^a, k^a \in \mathbb{Z}^6}\Tr\left[[A^I(-n^a-m^a-k^a),A^J(n^a)][A_I(m^a), A_J(k^a)]\right] \, ,\\
        V_2 &= -\lambda \sum_{n^a, m^a \in \mathbb{Z}^6}\Tr\left[2\pi(n^a+m^a)A^I(-n^a-m^a)[A_a(n^a), A_I(m^a)]\right] \, .
    \end{aligned}
\end{equation*}
Note that at least one of the modes summed over is a zero-mode. The interaction term with ghosts reads
\begin{equation*}
    V_{11}=-\lambda \sum_{m^a \in \mathbb{Z}^6}(2\pi m^a)\Tr \left[\bar{c}(m^a) [A_a(0), c(m^a)]\right] \, .
\end{equation*}
There are also interaction terms which only involve non-zero modes, but those are integrated over in the Wilsonian effective theory calculation, and since they do not depend on zero modes, they contribute to the effective action as a constant which can be ignored.

The interaction term for fermions is composed of a periodic and an anti-periodic component. These are completely disconnected from each other. We define $V_3,...,V_6$ to involve periodic spinors, given by 
\begin{equation*}
    \begin{aligned}
        V_3&=-\frac{i}{2}\lambda \sum_{r^a \in \mathbb{Z}^6}\Tr\left[\bar{\psi}^{\uparrow}_+(r^a)\tilde\Gamma^a[A_a(0), \psi^{\uparrow}_+(r^a)]\right]\,,\\
        V_4&= -\frac{i}{2}\lambda \sum_{r^a \in \mathbb{Z}^6}\Tr\left[\bar{\psi}^{\uparrow}_-(r^a)\tilde\Gamma^a[A_a(0), \psi^{\uparrow}_-(r^a)]\right]\,,\\
        V_5&=\frac{i}{4}\lambda \sum_{r^a \in \mathbb{Z}^6} \Tr \left[\bar{\psi}^{\uparrow}_+(r^a) \sigma^{\mu}_\uparrow[A_{\mu}(0), \psi^{\uparrow}_-(r_a)]\right]\,,\\
        V_6&=-\frac{i}{4}\lambda \sum_{r^a \in \mathbb{Z}^6} \Tr \left[\bar{\psi}^{\uparrow}_-(r^a) \bar{\sigma}^{\mu}_\uparrow[A_{\mu}(0), \psi^{\uparrow}_+(r_a)]\right] \, .
    \end{aligned}
\end{equation*}
The terms $V_7,..., V_{10}$ are obtained by swapping $\psi^{\uparrow}_{\pm} \leftrightarrow \psi^{\downarrow}_{\pm}$, $\sigma^\mu_\uparrow \leftrightarrow \sigma^\mu_\downarrow$, and considering half-integers winding modes. Here 
\begin{equation}
\begin{aligned}
    \sigma^{\mu}_\uparrow&=\begin{pmatrix}
            1 &1
        \end{pmatrix}\sigma^\mu \begin{pmatrix}
            1 \\
            1
        \end{pmatrix}=\sum_{i,j=1}^{2} \sigma^\mu_{ij}=2\begin{pmatrix}
            1 \\
            i\\
            0\\
            0
        \end{pmatrix}\\
       \sigma^{\mu}_\downarrow&=\begin{pmatrix}
            1 &-1
        \end{pmatrix}\sigma^\mu \begin{pmatrix}
            1 \\
            -1
        \end{pmatrix}=\sum_{ij=1}^{2}(-1)^{i+j}\sigma^\mu_{ij}=2\begin{pmatrix}
            1\\
            -i\\
            0\\
            0
        \end{pmatrix} \, .
        \end{aligned}
        \label{eq:SigmaComponents}
\end{equation}

We can now compute the effective action for zero modes derived from the interaction terms above. Starting with bosons and ghosts, for which the computation proceeds analogously to \cite{Laliberte_2024}, we have 
\begin{equation*}
    S_{0}^{eff} \supset \left \langle V_1 \right \rangle -\frac{1}{2} \left \langle V_2^2 \right \rangle -\frac{1}{2} \left \langle V_{11}^2 \right \rangle \, ,
\end{equation*}
and expectation values are computed using the propagators
\begin{equation*}
    \begin{aligned}
        \left \langle A^N(n^a) A^M(m^a) \right \rangle & =\frac{\delta^{MN} \delta_{n^a+m^a, 0}}{(2\pi n^a)^2} \, ,\\
        \left \langle \Bar{c}(n^a) c(m^a) \right \rangle & = \frac{\delta_{n^a,m^a}}{(2\pi n^a)^2} \, .
    \end{aligned}
\end{equation*}
The result reads 
\begin{equation*}
    \begin{aligned}
        &\left \langle V_1 \right \rangle = 9 \lambda^2 N S_{B_1} \Tr [A^I(0)A_I(0)] \\
        & \left \langle V_2^2 \right \rangle = 2 \lambda^2 N \left((17 S_{B_2}+S_{B_1})\Tr [A^a(0)A_a(0)]+S_{B_1} \Tr [A^\mu (0) A_\mu (0)] \right)\\
        & \left \langle V_{11}^2 \right \rangle =-2 \lambda^2 N S_{B_2} \Tr[A^a(0)A_a(0)] \, ,
    \end{aligned}
\end{equation*}
where the factors
\begin{equation}
    S_{B_1}=\sum_{n^a \in \mathbb{Z}^6} \nolimits' \frac{1}{(2\pi n^a)^2}, \quad S_{B_2}=\sum_{n^a \in \mathbb{Z}^6}\nolimits' \frac{(2\pi n^1)^2}{(2\pi n^a)^4} \, ,
    \label{eq:bosonicDiv}
\end{equation}
come from propagators in the loop. Note that these sums are divergent and have to be regularised, as discussed in \cite{Laliberte_2024}.

We now turn our attention to contributions to the effective action induced by integrating out fermions from the interaction terms $V_3,...V_{10}$. Fermionic propagators are inferred from (\ref{eq:Kinetic}) and read
\begin{equation*}
\left \langle \bar{\psi}^{u}_{p, \alpha}(r^a) \psi^v_{q, \beta}(s^a) \right \rangle= -i\frac{2\pi r^a \tilde\Gamma^a_{\alpha \beta}\delta^{u v}\delta_{p,q}\delta_{r^a, s^a}}{(2\pi r^a)^2} \, ,
\end{equation*}
where $u,v=\uparrow, \downarrow$, $p,q=\pm$ and $\alpha, \beta=1,...,8$ are $SO(6)$ spinor indices. It follows that the contribution of fermionic interactions to the zero-mode effective action is
\begin{equation}
    S_0^{eff} \supset -\frac{1}{2}\left \langle V_3^2 \right \rangle -\frac{1}{2} \left \langle V_4^2 \right \rangle -\left \langle V_5 V_6 \right \rangle -\frac{1}{2} \left \langle V_7^2 \right \rangle -\frac{1}{2} \left \langle V_8^2 \right \rangle-\left \langle V_9 V_{10} \right \rangle \,.
    \label{eq:fermeff}
\end{equation}
Recalling that $\psi^\uparrow_{\pm}$ has periodic boundary conditions and $\psi^\downarrow_{\pm}$ has anti-periodic ones, a straightforward application of Wick's theorem yields 
\begin{equation*}
    \begin{aligned}
        &\left \langle V_3^2 \right \rangle = -4 \lambda^2 N(2 S_{B_2}-S_{B_1})\Tr [A^a(0)A_a(0)] \\
        &\left \langle V_4^2 \right \rangle =  -4 \lambda^2 N(2 S_{B_2}-S_{B_1})\Tr [A^a(0)A_a(0)]\\
        & \left \langle V_7^2 \right \rangle =  -4 \lambda^2 N(2 S_{F_2}-S_{F_1})\Tr [A^a(0)A_a(0)]\\
        &\left \langle V_8^2 \right \rangle =  -4 \lambda^2 N(2 S_{F_2}-S_{F_1})\Tr [A^a(0)A_a(0)]\, ,
    \end{aligned}
\end{equation*}
where $S_{F_1}$ and $S_{F_2}$ read
\begin{equation*}
    S_{F_1}=\sum_{s^a \in \mathbb{L}_6} \frac{1}{(2\pi s^a)^2}, \quad S_{F_2}=\sum_{s^a \in \mathbb{L}_6} \frac{(2\pi s^1)^2}{(2\pi s^a)^4} \, .
\end{equation*}
These terms are also divergent but can be regularised with the same procedure as the bosonic ones, and the divergent part is the same \cite{Laliberte_2024}. In particular, the difference $S_{F_1}-S_{B_1}$ is a finite number which can be evaluated numerically. 

Cross terms in (\ref{eq:fermeff}) require a closer look. Let us analyse $\left \langle V_5 V_6 \right \rangle$. After contracting fermionic fields, we obtain
\begin{equation*}
    \left \langle V_5 V_6 \right \rangle = \lambda^2 N S_{B_1} \sigma^\mu _\uparrow \bar{\sigma}^\nu_\uparrow \Tr [A_\mu (0) A_\nu (0)] \, ,
\end{equation*}
which we can rearrange using (\ref{eq:SigmaComponents}) 
\begin{equation*}
    \sigma^\mu _\uparrow \bar{\sigma}^\nu_\uparrow \Tr [A_\mu (0) A_\nu (0)]=4\Tr[A^2_0(0)+A^2_1(0)] \, ,
\end{equation*}
and hence 
\begin{equation}
    \left \langle V_5 V_6 \right \rangle=4\lambda^2 N S_{B_1}\Tr[A^2_0(0)+A^2_1(0)] \, .
    \label{eq:V5V6}
\end{equation}
Analogously, one can compute 
\begin{equation}
    \left \langle V_9 V_{10} \right \rangle =4\lambda^2 N S_{F_1}\Tr[A^2_0(0)+A^2_1(0)] \, .
    \label{eq:V9V10}
\end{equation} 
Note that (\ref{eq:V5V6}) and (\ref{eq:V9V10}) imply that fermions in the loop only contribute to the mass of two of the components of bosonic matrices in the $SO(4)$ sector, thus breaking the $SO(4)$ symmetry. 

We can now summarise all contributions to the zero modes effective action (\ref{eq:ZeroEff}) to order $O(\lambda^2)$ in perturbation theory. First, the action $S_0$ for zero-modes at tree level receives contributions from both the bosonic part and the periodic components of fermions, and is given by
\begin{align*}
    S_0=-\frac{1}{2} \lambda^2 &\Tr \left[\frac{1}{2}[A^I(0),A^J(0)]^2+i\bar{\psi}_{+}^{\uparrow}(0)\tilde{\Gamma}^a[A_a(0), \psi_{+}^{\uparrow}(0)]+i\bar{\psi}_{-}^{\uparrow}(0)\tilde{\Gamma}^a[A_a(0), \psi_{-}^{\uparrow}(0)]\right.\\
    &\left. \vphantom{\frac{1}{2}}-i\bar{\psi}^{\uparrow}_+(0)  [A_0(0)+iA_1(0), \psi^{\uparrow}_{-}(0)]+i\bar{\psi}^{\uparrow}_ -(0)[A_0(0)-iA_1(0), \psi^{\uparrow}_{+}(0)]\right] \, .
\end{align*}
Including the above one-loop corrections, we obtain the full effective action
\begin{equation}
\begin{aligned}
    S_0^{eff}&=S_0-\lambda^2 N\left(4(S_{F_1}-S_{B_1})\Tr[A^2_0(0)+A^2_1(0)] \vphantom{\frac{8}{3}}\right.\\ 
    &\left.-8S_{B_1}\Tr[A^2_2(0)+A^2_3(0)] +\frac{8}{3}(S_{F_1}-S_{B_1})\Tr[A^2_a(0)]\right) \, .
    \label{eq:Interchanging_Eff}
\end{aligned}
\end{equation}
Note that the choice of interchanging boundary conditions (\ref{eq:reversedBC}) breaks the $SO(4)$ symmetry subgroup in two ways: through the interaction term, which couples fermionic zero-modes to specific components of the bosonic field; and through the mass matrix, which is diagonal but takes different values for different components in the four-dimensional subspace. Moreover, these boundary conditions effectively reduce the number of fermionic degrees of freedom which contribute to loop diagrams, due to the decoupling of $\psi^\uparrow$ and $\psi^\downarrow$ modes. As a consequence, cancellation of divergences at the one-loop level fails. In particular, we find a divergent mass 
\begin{equation*}
    M^2_2=M^2_3=-16 \lambda^2NS_{B_1}
\end{equation*}
for two components of the bosonic field. This highlights a complication in generating a mass term alongside fermionic degrees of freedom from torus compactifications of IKKT with mixed boundary conditions. We will show in the next subsection that this issue also arises with alternative choices of boundary conditions. As we will see in section \ref{sec:BFSS_Mixed}, the same issue does not arise in BFSS theory, where we explicitly find a class of mixed boundary conditions which does not affect the cancellation at one-loop level and gives a finite result. This constitutes an interesting qualitative difference between the two models, which would be worth exploring further.

We can consider renormalising the divergence by introducing an \emph{ad hoc} mass counterterm in the compact action on the torus (\ref{eq:Sint}). This is artificial from the point of view of the original IKKT model, where this term is forbidden by supersymmetry and global $SO(10)$. However, we explicitly showed how both these symmetries are broken in the compact action, so this term is allowed at the level of the EFT. We have freedom in choosing the renormalisation scheme. The choice we make is to cancel the divergence by introducing the counterterm action
\begin{equation*}
    S_{CT}=-\frac{\lambda^2}{2} \int \frac{d^6 \sigma}{(2\pi L)^{-6}}16NS_{F_1} \Tr[A^2_2(0)+A^2_3(0)] \, ,
\end{equation*}
such that 
\begin{multline*}
    S^{eff}_0=S_0-\ln \left \langle e^{-S_{int}}\right \rangle-\ln Z_{kin}+S_{CT}\\
    =S_0-\lambda^2 N\Biggl(4(S_{F_1}-S_{B_1})\Tr[A^2_0(0)+A^2_1(0)]\\ 
    +8(S_{F_1}-S_{B_1})\Tr[A^2_2(0)+A^2_3(0)] +\frac{8}{3}(S_{F_1}-S_{B_1})\Tr[A^2_a(0)]\Biggr) \, .
\end{multline*}
This choice is natural in the sense that it reinserts the contribution coming from fermions in the loop, which have been lost due to the mixed boundary conditions. Of course, other choices are possible, and the finite value of the resulting mass is scheme-dependent. We then rescale fields back by inverting (\ref{eq:fieldsRescaling}), and Wick rotate to restore Lorentzian signature, resulting in
\begin{multline*}
    S_0=-\frac{1}{2 g^2_{eff}} \Tr \left[\frac{1}{2}[A^I(0),A^J(0)]^2+\bar{\psi}_{+}^{\uparrow}(0)\tilde{\Gamma}^a[A_a(0), \psi_{+}^{\uparrow}(0)]+\bar{\psi}_{-}^{\uparrow}(0)\tilde{\Gamma}^a[A_a(0), \psi_{-}^{\uparrow}(0)]\right. \\ \left.
    \vphantom{\frac{1}{2}}-\bar{\psi}^{\uparrow}_+(0)  [A_0(0)-A_1(0), \psi^{\uparrow}_{-}(0)]+\bar{\psi}^{\uparrow}_ -(0)[A_0(0)+A_1(0), \psi^{\uparrow}_{+}(0)]\right]\,,
\end{multline*}
and 
\begin{multline*}
    S_0^{eff}=S_0+\frac{N}{L^2}\left(4(S_{F_1}-S_{B_1})\Tr[A^2_0(0)-A^2_1(0)]\vphantom{\frac{1}{2}}\right. \\ \left.
    -8(S_{F_1}-S_{B_1})\Tr[A^2_2(0)+A^2_3(0)] -\frac{8}{3}(S_{F_1}-S_{B_1})\Tr[A^2_a(0)]\right) \, .
\end{multline*}
As expected, the theory we obtain shows mixed qualitative features compared to the purely periodic or anti-periodic cases: here, parts of the fermions are still present in the zero modes action, alongside a mass term. Both contributions break Lorentz symmetry explicitly.

Note that there are other options for boundary conditions, which also produce infinite mass terms in the effective action and require renormalisation. For example, we can give opposite boundary conditions to different components of the two-dimensional spinor or duplicate fermions and give each component different boundary conditions. 

Another possibility would be to consider the theory to be compactified on $\mathbb{T}^4$. Then, upon considering the decompactification limit with mixed boundary conditions, we can obtain an effective theory with a broken $SO(6)$ subgroup and an unbroken $SO(4)$.
\subsection{Chiral Reversing boundary conditions}
A further type of boundary condition that can be considered is swapping chiral components of fermions when cycling around the torus. These are implemented by
\begin{equation*}
    (U^a)^{-1} \psi_+^A U^a =\psi_-^A, \qquad (U^a)^{-1} \psi_-^A U^a =\psi_+^A \, .
\end{equation*}
In other words, we decompose
\begin{equation}
\label{eq:ChiralReverse}
    \psi^A_+=\frac{1}{\sqrt{2}}(\psi_\uparrow^A+\psi^A_\downarrow), \qquad \psi^A_-=\frac{1}{\sqrt{2}}(\psi_\uparrow^A-\psi^A_\downarrow) \, ,
\end{equation}
where $\psi_\uparrow^A$ has periodic boundary conditions and $\psi_\downarrow^A$ is anti-periodic, and their Hilbert space decomposition (\ref{eq:TensorHilbert}) reads
\begin{equation*}
    \begin{aligned}
        \psi_{\uparrow}^{A}&=\sum_{n^b \in \mathbb{Z}^6}\psi_{\uparrow}^{A}(n^b) \otimes e^{in^b p^b} \\
        \psi_{\downarrow}^{A}&=\sum_{r^b \in \mathbb{L}_6}\psi_{\downarrow}^{A}(r^b) \otimes e^{ir^b p^b} \, \, .
    \end{aligned}
\end{equation*}
Substituting (\ref{eq:ChiralReverse}) into the compact action (\ref{eq:CompactAction}), we get the fermionic action
\begin{multline*}
    S_F^{CR}=\frac{1}{2 g^2_{eff}}\int \frac{d^6 \sigma}{(2\pi L)^{-6}} \Tr\left[\bar{\psi}_{\uparrow}^{A}\tilde{\Gamma}^a\partial_a \psi_{\uparrow}^{A}+\bar{\psi}_{\downarrow}^{A}\tilde{\Gamma}^a\partial_a \psi_{\downarrow}^{A}
    -i\bar{\psi}_{\uparrow}^{A}\tilde{\Gamma}^a[A_a, \psi_{\uparrow}^{A}]-i\bar{\psi}_{\downarrow}^{A}\tilde{\Gamma}^a[A_a,\psi_{\downarrow}^{A}]  \right. \\ \left.
    +i\bar{\psi}^{A}_\uparrow (\sigma^j)_{AB} [A_j, \psi^{B}_{\uparrow}]-i\bar{\psi}^{A}_\downarrow (\sigma^j)_{AB} [A_j, \psi^{B}_{\downarrow}]\right]\,,
\end{multline*}
where the index $j=1,2,3$ runs over the three space components of $SO(4)$ and the superscript $CR$ stands for \textit{chiral reversing} boundary conditions. The dependence on the momentum variable \(\sigma\) is obtained through the mode expansion

\begin{equation*}
    \begin{aligned}
        \psi_{\uparrow}^{A}(\sigma)&=\sum_{n^b \in \mathbb{Z}^6}\psi_{\uparrow}^{A}(n^b) e^{2\pi iLn^b \sigma^b} \\
        \psi_{\downarrow}^{A}(\sigma)&=\sum_{r^b \in \mathbb{L}_6}\psi_{\downarrow}^{A}(r^b) e^{2\pi iLr^b \sigma^b} \, \, .
    \end{aligned}
\end{equation*}
Note that the $A_0$ component of the bosonic vector is not coupled to spinors, due to the relation (\ref{eq:traceproperty}) which forbids terms of the form $\bar{\psi}^{A}_\downarrow (\sigma^0)_{AB} [A_0, \psi^{B}_{\uparrow}]$. 
The fermionic propagator in the space of winding modes is
\begin{equation*}
    \left \langle \bar{\psi}^{A}_{u, \alpha}(r^a) \psi^B_{v, \beta}(s^a) \right \rangle= -i\frac{2\pi r^a \tilde\Gamma^a_{\alpha \beta}\delta^{A,B}\delta_{u,v}\delta_{r^a, s^a}}{(2\pi r^a)^2} \, ,
\end{equation*}
where $u,v=\uparrow, \downarrow$. Integrating out the non-zero fermionic modes, we get in this case (after the usual field redefinition (\ref{eq:fieldsRescaling}))
\begin{equation}
    S_0^{eff}=S_0^{CR}-\lambda^2 N\left(-8S_{B_1}\Tr[A^2_0(0)]+4(S_{F_1}-S_{B_1})\Tr[A^2_i(0)]+\frac{8}{3}(S_{F_1}-S_{B_1})\Tr[A^2_a(0)]\right) \, ,
    \label{eq:effCR}
\end{equation}
and 
\begin{equation*}
    S^{CR}_0=-\frac{1}{2} \lambda^2 \Tr \left[\frac{1}{2}[A^I(0),A^J(0)]^2+i\bar{\psi}_{\uparrow}^{A}(0)\tilde{\Gamma}^a[A_a(0), \psi_{\uparrow}^{A}(0)]
    +i\bar{\psi}^{A}_\uparrow(0) (\sigma^i)_{AB} [A_i(0), \psi^{B}_{\uparrow}(0)]\right] \, .
\end{equation*}
Note that this effective theory for the zero-modes 
is significantly different from (\ref{eq:Interchanging_Eff}). In particular, it features an explicit coupling of fermionic modes to all components of the bosonic vectors, except the zeroth one. The latter precisely corresponds to the bosonic component, which acquires a divergent mass, due to the cancellation of some of the fermionic interactions which contribute at one loop. We can add a counterterm to the compact action in order to renormalise this divergence, and our choice of renormalisation scheme at one loop is
\begin{equation*}
    S_{CT}=-\frac{\lambda^2}{2} \int \frac{d^6 \sigma}{(2\pi L)^{-6}}16N\,S_{F_1} \Tr[A^2_0(0)] \, ,
\end{equation*}
such that the renormalised effective action for zero modes reads
\begin{multline*}
    S_0^{eff,R}=S_0^{eff}+S_{CT}=S_0^{CR}-\lambda^2 N\left(8(S_{F_1}-S_{B_1})\Tr[A^2_0(0)]\vphantom{\frac{1}{2}}\right. \\ \left.
    +4(S_{F_1}-S_{B_1})\Tr[A^2_i(0)]+\frac{8}{3}(S_{F_1}-S_{B_1})\Tr[A^2_a(0)]\right) \, ,
\end{multline*}
which is finite and preserves the $SO(3)$ symmetry subgroup of the $SO(4)$. We can rescale fields back and go to Lorentzian signature 
\begin{multline*}
    S_0^{eff,R}=S_0^{eff}+S_{CT}=S_0^{CR}+\frac{N}{L^2}\left(8(S_{F_1}-S_{B_1})\Tr[A^2_0(0)]\vphantom{\frac{1}{2}}\right. \\ \left.
    -4(S_{F_1}-S_{B_1})\Tr[A^2_i(0)]-\frac{8}{3}(S_{F_1}-S_{B_1})\Tr[A^2_a(0)]\right) \, .
\end{multline*}
The conclusion is that mixed boundary conditions for fermions in IKKT theory dramatically change the zero-mode effective theory. They cause divergent contributions to mass terms to appear in the effective action. These can be renormalised in order to obtain a finite result. As a result, a subgroup of $SO(1,3) \times SO(6)$ is broken by the newly generated mass terms.

In the next two sections, we will turn our attention to BFSS theory. We will see that, in this case, the situation is different and, in particular, there is a class of mixed boundary conditions which do not break any subgroup of the global symmetry $SO(3) \times SO(6)$.
\section{Compactification of BFSS theory}
\label{sec:BFSSComp}
In this section, we extend the mechanism of torus compactification to the BFSS matrix theory. We will consider the cases of periodic and anti-periodic boundary conditions for fermions, and show that the result is equivalent to the corresponding cases in IKKT theory.

In BFSS theory, compactifications on a six-dimensional torus are ill-defined \cite{Seiberg:1997, Sen:1997}. We therefore consider compactifications of the complementary coordinates on a three-dimensional square torus $\mathbb{T}^3$, where each of the dimensions has size $2 \pi L$. We expect that, in the decompactification limit, the results should be qualitatively equivalent, still producing different mass terms for the two sets of directions. We will show that this is indeed the case. 

BFSS matrix theory is an $SU(\mathcal{N})$ gauge theory with gauge field $A$ in the adjoint representation. The other degrees of freedom are a nine-component vector $X^I$, $I=1,...,9$ of bosonic $\mathcal{N} \times \mathcal{N}$ matrices and a sixteen-component spinor $\psi_\alpha$, $\alpha=1,..., 16$, where each component is a Grassmann-valued $\mathcal{N} \times \mathcal{N}$ matrix. It is a quantum mechanical theory, with time variable $t$, and action, in Lorentzian signature, given by
\begin{equation*}
    S_{BFSS}=\int dt \frac{1}{g^2} \Tr\left[\frac{1}{2}(\mathcal{D}_tX^I)^2+\frac{1}{4}[X^I,X^J]^2+i \bar{\psi} \mathcal{D}_t\psi-\bar{\psi}\Gamma^I[X_I, \psi]\right] \, ,
\end{equation*}
where $\Gamma^I$ are gamma matrices in nine dimensions. This theory is maximally supersymmetric and has a global $SO(9)$ symmetry, which can be interpreted as a subgroup of target space Lorentz symmetry, where the $X^I$ matrices play the role of space coordinates. It has been conjectured \cite{Banks_1996} to be dual to M-theory in the Infinite Momentum Frame and, as such, it provides a non-perturbative definition, which makes it a very interesting candidate for a theory of quantum gravity. Analysing compactifications of matrix theory thus provides a potential path to a quantum gravity theory in the symmetry-broken phase, where three large space dimensions emerge. 
In this section, we will analyse the cases of anti-periodic and periodic boundary conditions for fermions, and in the next section extend the discussion to include mixed boundary conditions.

It will be convenient to work in Euclidean signature. To this end, we Wick rotate by means of
\begin{equation*}
    t \rightarrow it, \quad A \rightarrow iA, \quad \Gamma^I \rightarrow i\Gamma^I, \quad \psi \rightarrow i\psi\, ,
\end{equation*}
and we obtain the Euclidean action
\begin{equation}
    S_{BFSS}^E=\int dt \frac{1}{g^2} \Tr\left[\frac{1}{2}(\mathcal{D}_tX^I)^2-\frac{1}{4}[X^I,X^J]^2+ \Bar{\psi} \mathcal{D}_t\psi-i\Bar{\psi}\Gamma^I[X_I, \psi]\right] \, .
    \label{eq:EuclideanBFSS}
\end{equation}
Note that, in this signature, gamma matrices satisfy
\begin{equation*}
    \{\Gamma^I, \Gamma^J\}=-2\delta^{IJ} \, ,
\end{equation*}
with a minus sign. The next step is to define a translation operator on the torus $\mathbb{T}^3$. This is done analogously to the method of section \ref{sec:ReviewIKKT}, by defining a Hilbert space acted upon by matrices, which is decomposed into two components, as in equation (\ref{eq:Toepldecomposition}). It follows that the degrees of freedom of the theory are decomposed as
\begin{equation}
    \begin{aligned}
        &A(t)= \sum_{n^b \in \mathbb{Z}^3} A(t, n^b) \otimes e^{i n^b p^b} \, , \\
        &X^\mu(t)=\sum_{n^b \in \mathbb{Z}^3} X^\mu(t, n^b)\otimes e^{i n^b p^b} \, ,\\
        &X^a(t)=\sum_{n^b \in \mathbb{Z}^3} X^a(t, n^b)\otimes e^{i n^b p^b}+2\pi L \mathbb{I}_N \otimes q^a \, ,\\
        &\psi(t)=\sum_{r^b \in \mathbb{L}_3} \psi(t, r^b) \otimes e^{i r^b p^b} \, ,
    \end{aligned}
    \label{eq:BFSSDecomp}
\end{equation}
where we are considering the case of anti-periodic fermions explicitly, and we will comment on the periodic case later. The lattice \(\mathbb{L}_3\) is the three-dimensional analogue of (\ref{eq:shitlattice}), namely
\begin{equation}
    \label{eq:L3}
\mathbb{L}_3=\left(\mathbb{Z}+\frac{1}{2}\right) \times \left(\mathbb{Z}+\frac{1}{2}\right) \times \left(\mathbb{Z}+\frac{1}{2}\right) \, ,
\end{equation}
where lattice sites represent half-integer winding modes in each direction. Here, indices are $\mu=4,...,9$ and $a=1,2,3$, such that indices labelled by Latin letters $a,b,..$ correspond to compact directions (note that this is different from previous sections). Note that all the degrees of freedom now have an explicit time dependence. We substitute the decomposition (\ref{eq:BFSSDecomp}) into the Euclidean action (\ref{eq:EuclideanBFSS}) and define the mode expansion of the fields 
\begin{equation*}
    \begin{aligned}
        &A(t, \sigma)= \sum_{n^b \in \mathbb{Z}^3} A(t, n^b) e^{2 \pi iL n^b \sigma^b} \, , \\
        &X^\mu(t, \sigma)=\sum_{n^b \in \mathbb{Z}^3} X^\mu(t, n^b)e^{2 \pi iL n^b \sigma^b}\, ,\\
        &X^a(t, \sigma)=\sum_{n^b \in \mathbb{Z}^3} X^a(t, n^b)e^{2 \pi iL n^b \sigma^b}\, ,\\
        &\psi(t, \sigma)=\sum_{r^b \in \mathbb{L}_3} \psi(t, r^b) e^{2 \pi iL r^b \sigma^b}\, ,\\
        &c(t,\sigma)= \sum_{n^b \in \mathbb{Z}^3} c(t, n^b) e^{2 \pi iL n^b \sigma^b} \, ,
    \end{aligned}
\end{equation*}
where $\sigma \in [0, L^{-1}]^3$ is the momentum mode along the torus. We obtain the compact BFSS action
\begin{multline}
    S^C_{BFSS}=\frac{1}{g^2_{eff}} \int\frac{d^3 \sigma dt}{(2\pi L)^{-3}} \Tr \left[\frac{1}{4}F^{AB}F_{AB}+\frac{1}{2}\mathcal{D}_AX_\mu \mathcal{D}^A X^\mu \right. \\ \left.
    -\frac{1}{4}[X^\mu, X^\nu]^2+\bar{\psi} \mathcal{D}_t \psi+\bar{\psi}\Gamma^a\mathcal{D}_a\psi-i\bar{\psi}\Gamma^\mu [X_\mu, \psi]\right]\, ,
    \label{eq:BFSSCompact}
\end{multline}
where, for convenience, we have defined an index $A=0, 1,2,3$, corresponding to the time and space coordinates of the fields. The effective gauge coupling is defined as $g_{eff}^2=\frac{g^2}{M}$, $M$ being the dimension of the Hilbert subspace on which $p^a$ and $q^a$ are defined. We choose to work in Lorentz gauge $\partial_B X^B=0$, where $X^0(t, \sigma)=A(t, \sigma)$, implying that we must add a ghost term to the compact action 
(\ref{eq:BFSSCompact})
\begin{equation*}
    S_{gh}=\frac{1}{g_{eff}^2} \int \frac{d^3 \sigma\, dt}{(2 \pi L)^{-3}}\Tr[ \partial^A \bar{c} \mathcal{D}_A c] \, .
\end{equation*}
Furthermore, we Fourier transform the time variable in order to work in energy space, and perform the usual field rescaling, where we also rescale the time variable to make it dimensionless
\begin{equation}
    X^\alpha \rightarrow \lambda L X^{\alpha}, \quad \psi \rightarrow \lambda L^{3/2} \psi, \quad c \rightarrow \lambda L c,\quad  t\rightarrow \frac{t}{L}, \quad \lambda^2 =\frac{g^2_{eff}}{L^3} \, ,
    \label{eq:BFSSrescaling}
\end{equation}
where we introduced the index $\alpha=0,...,9$ and $X^0=A$. As a result, we obtain the kinetic action
\begin{multline}
    S_{kin}=\int dp \,\Tr\left[\sum_{n^a \in \mathbb{Z}^3}\frac{1}{2}(p^2+n_an^a)X_\alpha(p, n^a) X^\alpha(-p, -n^a)\right. \\ \left.
    +\sum_{s^a \in \mathbb{L}_3 }i(p+\Gamma^as_a )\bar{\psi}(p, s^a)\psi(p, s^a) +\sum_{n^a \in \mathbb{Z}^3}(p^2+n_an^a)\bar{c}(p, n^a) c(p, n^a)\right] \, ,
    \label{eq:BFSSKinetc}
\end{multline}
and we observe that non-zero modes are massive. In particular, in the regime $L \gg g_{eff}^{2/3}$, the parameter $\lambda$ is perturbative, and we can integrate these modes out at one-loop order. Analogously to the discussion around (\ref{eq:ZeroEff}), we follow a Wilsonian approach to obtain an effective action for zero modes. The relevant interaction terms are those which couple zero modes to non-zero modes. The purely bosonic ones read 
\begin{multline*}
        V_1=-\frac{\lambda^2}{4}\tilde{\sum_{n^a, m^a,l^a \in \mathbb{Z}^3}} \int dp dq dr \Tr\left[[X^\alpha(p, n^a), X^\beta(q, m^a)]\right. \\ \left.[X_\alpha (r, l^a), X_\beta(-p-q-r, -n^a-m^a-l^a)]\vphantom{X^{\beta}} \right] \, ,
\end{multline*}
\begin{multline*}
    V_2=-\lambda \Tilde{\sum_{n^a, m^a \in \mathbb{Z}^3}} \int dpdq\Tr\left[(p+q)X_\alpha(-p-q,-n^a-m^a)[A(p,n^a), X^\alpha(q,m^a)]\right. \\ \left.
    +(n^a+m^a)X_\alpha(-p-q,-n^a-m^a)[X_a(p,n^a), X^\alpha(q,m^a)]\right] \, ,
\end{multline*}
where $\Tilde{\sum}$ means that, in each term of the sum, one of the winding numbers is set to zero. Interactions involving fermions and ghost fields read
\begin{multline*}
    V_3=-i\lambda \sum_{s^a \in \mathbb{L}_3} \int dp dq \Tr\left[\bar{\psi}(p+q, s^a)[A(p, 0), \psi(q, s^a)]\right. \\ \left.
    +\bar{\psi}(p+q, s^a)\Gamma^a[X_a(p,0), \psi(q, s^a)]\right] \, ,
\end{multline*}
\begin{equation*}
    V_4=-i\lambda \sum_{s^a \in \mathbb{L}_3} \int dp dq \Tr \left[\bar{\psi}(p+q, s^a)\Gamma^\mu[X_\mu(p,0), \psi(q, s^a)]\right] \, ,
\end{equation*}
\begin{multline*}
    V_5=\lambda \sum_{n^a\in\mathbb{Z}^3} \int dp dq \Tr \left[(p+q)\bar{c}(p+q, n^a)[A(p, 0), c(q, n^a)]\right. \\ \left.
    +n^a\bar{c}(p+q, n^a)[X_a(p, 0), c(q, n^a)]\right] \, ,
\end{multline*}
and we note that, in this case, it is convenient not to decompose spinors, for the sake of simplicity. In terms of the interaction potentials, the zero-mode effective action is given by
\begin{equation*}
    S_0^{eff}=S_0-\ln \left \langle e^{-S_{int}}\right \rangle=S_0+\left \langle V_1 \right \rangle-\frac{1}{2}\sum_{i=2}^5\left \langle V_i^2\right \rangle
\end{equation*}
and expectation values are computed using Wick's theorem, where building blocks are two-point functions derived from the kinetic action (\ref{eq:BFSSKinetc})
\begin{equation*}
    \begin{aligned}
        &\left \langle X_\alpha(p, n^a)X_\beta(q, m^a) \right \rangle =\frac{\delta_{\alpha \beta }\delta(p+q)\delta_{n^a+m^a,0}}{p^2+n^an_a} \, ,\\
        &\left \langle \bar{\psi}(p, s^a)\psi(q, r^a)\right \rangle = -i\frac{\delta(p-q)\delta_{r^a, s^a}(p-\Gamma^a s_a)}{p^2+s^a s_a} \, ,\\
        &\left \langle \bar{c}(n^a,p)c(m^a, q)\right \rangle=\frac{\delta(p-q)\delta_{n^a,m^a}}{p^2+n^a n_a} \,.
    \end{aligned}
\end{equation*}
As a result, the averages read
\begin{equation}
    \begin{aligned}
        &\left \langle V_1\right \rangle =9 \lambda^2 N S_{B_1} \int dp \Tr[X^\alpha(-p,0)X_\alpha(p,0)] \, ,\\
        &\left \langle V_2^2 \right \rangle =2\lambda^2 N \int dp \left[(17 S_{B_3}+S_{B_1})\Tr[A(-p,0)A(p,0)] 
        \right. \\ 
        &\left.+(17S_{B_2}+S_{B_1})\Tr[X^a(-p,0)X_a(p,0)]+S_{B_1}\Tr [X^\mu(-p,0)X_\mu(p,0)]\right] \, ,\\
        &\left \langle V_3^2\right \rangle=-16\lambda^2 N \int dp \left[(2S_{F_3}-S_{F_1})\Tr[A(-p,0)A(p,0)]+(2S_{F_2}-S_{F_1})\Tr[X^a(-p,0)X_a(p,0)]\right]\, , \\
        &\left \langle V_4^2\right \rangle=16\lambda^2N \int dp S_{F_1} \Tr[X^\mu(-p,0)X_\mu(p,0)] \, , \\
        &\left \langle V_5^2\right \rangle=-2 \lambda^2 N \int dp \left[S_{B_3}\Tr[A(-p,0)A(p,0)]+S_{B_2}\Tr[X^a(-p,0)X_a(p,0)]\right] \, ,
    \end{aligned}
    \label{eq:BFSSexpect}
\end{equation}
where propagators in the loop give rise to factors
\begin{equation}
\label{eq:BFSSFactors}
    \begin{aligned}
        &S_{B_1}=\sum_{n^a\in \mathbb{Z}^3}\nolimits' \int_{-\infty}^\infty dp \frac{1}{p^2+n^an_a} \, , \qquad S_{F_1}=\sum_{r^a\in \mathbb{L}_3} \int_{-\infty}^\infty dp \frac{1}{p^2+r^ar_a} \, ,\\
        &S_{B_2}=\sum_{n^a\in \mathbb{Z}^3}\nolimits' \int_{-\infty}^\infty dp \frac{(n^1)^2}{(p^2+n^an_a)^2}\, , \qquad S_{F_2}=\sum_{r^a\in \mathbb{L}_3} \int_{-\infty}^\infty dp \frac{(r^1)^2}{(p^2+r^ar_a)^2} \, ,\\
        &S_{B_3}=\sum_{n^a\in \mathbb{Z}^3}\nolimits' \int_{-\infty}^\infty dp \frac{p^2}{(p^2+n^an_a)^2} \, , \qquad S_{F_3}=\sum_{r^a\in \mathbb{L}_3} \int_{-\infty}^\infty dp \frac{p^2}{(p^2+r^ar_a)^2}\, ,
    \end{aligned}
\end{equation}
and primed sums run over non-zero modes only. These factors are infinite, but can be regularised, generalising the procedure of \cite{Laliberte_2024}. We will elaborate on this in the Appendix \ref{sec:regularise}. We will also show that bosonic and fermionic terms have the same infinite part, which therefore cancels out when considering differences. 

We collect all terms (\ref{eq:BFSSexpect}) together, invert the rescaling (\ref{eq:BFSSrescaling}), and Wick rotate back to Lorentzian signature to obtain, for the zero modes effective action,
\begin{multline*}
    S_0^{eff}=S_0-\frac{1}{2}\int dt \left[M^2_A \Tr[A(t,0)A(t,0)]\vphantom{X^a}\right. \\ \left.
    -M^2_{X^\mu}\Tr[X_\mu (t,0)X^\mu(t,0)]-M^2_{X^a}\Tr[X_a(t,0) X^a(t,0)]\right] \,.
\end{multline*}
Here, the action $S_0$ reads
\begin{equation*}
    S_0=\frac{1}{g^2_{eff}}\int dt \Tr\left[\frac{1}{2}\mathcal{D}_tX_I(t,0)\mathcal{D}_tX^I(t,0)+\frac{1}{4}[X^I(t,0), X^J(t,0)]^2+i \partial_t\bar{c}(t,0) \mathcal{D}_tc(t,0)\right] \, ,
\end{equation*}
where $I,J=1,...,9$ and the masses are 
\begin{equation*}
    \begin{aligned}
        &M^2_A=16\frac{N}{L^2}\left(S_{F_1}-S_{B_1}+2(S_{B_3}-S_{F_3})\right) \,,\\
        &M^2_{X^\mu}=16\frac{N}{L^2}(S_{F_1}-S_{B_1})\, , \\
        &M^2_{X^a}=16 \frac{N}{L^2}\left(S_{F_1}-S_{B_1}+2(S_{B_2}-S_{F_2})\right) \, .
    \end{aligned}
\end{equation*}
By evaluating the $p$ integrals in (\ref{eq:BFSSFactors}), we find that 
\begin{equation*}
    S_{B_1}=2S_{B_3}, \qquad S_{F_1}=2S_{F_3} \, .
\end{equation*}
Moreover, these factors are related by the equations
\begin{equation*}
    S_{B_1}= S_{B_3}+3S_{B_2}, \qquad S_{F_1}= S_{F_3}+3S_{F_2} \, .
\end{equation*}
Altogether,  the mass terms then simplify to
\begin{equation}
    M^2_A=0, \quad M^2_{X^\mu}= 16\frac{N}{L^2}(S_{F_1}-S_{B_1}), \quad M^2_{X^a}=\frac{32}{3} \frac{N}{L^2}(S_{F_1}-S_{B_1}) \, ,
    \label{eq:BFSSMasses}
\end{equation}
where the difference $(S_{F_1}-S_{B_1})$ is finite after regularization, as shown in the Appendix \ref{sec:regularise}. The mass terms (\ref{eq:BFSSMasses}) show that, upon compactification on $\mathbb{T}^3$, with anti-periodic boundary conditions for fermions, the effective theory has a global symmetry $SO(3) \times SO(6)$, thus breaking the initial $SO(9)$. This extends the result of \cite{Laliberte_2024} to BFSS theory, proving that a mass term can be generated by compactification. It also hints towards the possibility that the BFSS model enjoys symmetry breaking analogously to what is observed in numerical studies of the IKKT model. Furthermore, the gauge field remains massless in the effective theory. Note that this is not due to a supersymmetric cancellation, which does not occur in the case of anti-periodic boundary conditions, but to the fact that bosonic and fermionic terms cancel out separately due to gauge invariance. 

The case of periodic boundary conditions for fermions is obtained straightforwardly by noting that, in that case, $S_{F_1}=S_{B_1}$. It follows that
\begin{equation*}
    M^2_A=M^2_{X^\mu}=M^2_{X^a}=0 \, ,
\end{equation*}
hence no mass term is generated. The zero-mode effective action now reads
\begin{multline*}
    S_0^{eff}=S_0^P=\int dt \frac{1}{g^2_{eff}} \Tr\left[\frac{1}{2}(\mathcal{D}_tX^I(t,0))^2
    +\frac{1}{4}[X^I(t,0),X^J(t,0)]^2\right. \\ \left.+i \bar{\psi}(t,0) \mathcal{D}_t\psi(t,0)-\bar{\psi}(t,0)\Gamma^I[X_I(t,0), \psi(t,0)]+i \partial_t\bar{c}(t,0) \mathcal{D}_tc(t,0)\vphantom{\frac{1}{2}}\right] \, ,
\end{multline*}
which is the same as the original BFSS action with an effective coupling constant.
\section{BFSS with Mixed Boundary Conditions}
\label{sec:BFSS_Mixed}
In this section, we consider a class of mixed boundary conditions for fermions which give rise to an effective theory with intermediate qualitative features. We will show that we can obtain a zero-mode effective theory where half the fermions are quenched, the remaining half being still present, and a mass term is generated. 

In contrast to the IKKT case treated above, we will show that BFSS theory admits a class of mixed boundary conditions which preserves the divergence cancellation at one loop. As a consequence, the effective action is finite and does not require renormalisation. Most importantly, we will show that it features both a symmetry-breaking mass term and fermionic fields.

We consider the decomposition of sixteen-component fermionic degrees of freedom introduced in \cite{Kim2003}, which we review here. We choose the $\Gamma^I$ matrices, $I=1,...,9$, to be decomposed such that
\begin{equation*}
    \Gamma^a =\begin{pmatrix}
        -\sigma^a \otimes \mathbb{I}_4 & 0\\
        0 & \sigma ^a \otimes \mathbb{I}_4
    \end{pmatrix}, \qquad \Gamma^\mu=\begin{pmatrix}
        0 & \mathbb{I}_2 \otimes \gamma^\mu\\
       - \mathbb{I}_2 \otimes \gamma^\mu & 0
    \end{pmatrix} \, ,
\end{equation*}
where $\gamma_\mu=-\gamma_\mu^\dagger$ are anti-Hermitian $4 \times 4$ matrices satisfying the Clifford algebra relation
\begin{equation*}
    \{\gamma^\mu, \gamma^\nu\}=-2 \delta^{\mu \nu} \mathbb{I}_4 \, .
\end{equation*}
The charge conjugation matrix is 
\begin{equation*}
    C_9= \begin{pmatrix}
        0 & -i\sigma_2 \otimes \mathbb{I}_4\\
        i \sigma_2 \otimes \mathbb{I}_4 & 0
    \end{pmatrix} \, ,
\end{equation*}
which allows for spinor decomposition in chiral components\footnote{Note that we inverted the notation of \cite{Kim2003} for spinor indices, in order to match with our conventions in previous sections.} 
\begin{equation*}
    \psi=\begin{pmatrix}
    \lambda^{A \alpha } \\
    i (\sigma^2)^{AB}\lambda^{*}_{B\alpha}
    \end{pmatrix}, \qquad A=1,2\,, \quad \alpha=1,...,4 \, \,.
\end{equation*}
It follows that the fermionic part of the compact BFSS Lagrangian (\ref{eq:BFSSCompact}) becomes
\begin{equation}
    g^2 \mathcal{L}_{ferm}=\Tr\left[2\lambda^\dagger_\alpha \mathcal{D}_t \lambda^\alpha -2i\lambda^\dagger_\alpha \sigma^a \mathcal{D}_a \lambda^{\alpha}-i \lambda^\dagger_\alpha i\sigma^2(\gamma^\mu)_{\alpha \beta}  [X_\mu, \lambda^*_\beta]-i \lambda^{T,\alpha} i\sigma^2(\gamma^\mu)^{\alpha \beta } [X_\mu, \lambda_{\beta}]\right] \, ,
    \label{eq:BFSS_Mixed_Lag}
\end{equation}
where we only wrote down the $\alpha$ components of spinor indices explicitly. One notable feature of the Lagrangian (\ref{eq:BFSS_Mixed_Lag}) is its invariance under 
\begin{equation*}
    \lambda \rightarrow \lambda^{*}, \qquad \lambda^{*} \rightarrow \lambda \, ,
\end{equation*}
which can be checked explicitly from the above expression (for the first two terms, transposition gives a minus sign due to the Grassmann nature of variables, and a further minus sign comes from taking the derivative by parts). For this reason, we consider boundary conditions which implement this transformation when wrapping around a closed loop in any of the torus directions, namely
\begin{equation*}
    (U^a)^{-1} \lambda^\alpha U^a =\lambda_\alpha^{*}, \qquad (U^a)^{-1} \lambda_\alpha^{*} U^a =\lambda^\alpha \, .
\end{equation*}
This implies that we can decompose spinors into periodic and anti-periodic components
\begin{equation*}
        \lambda=\frac{1}{\sqrt{2}}(\lambda^\uparrow+i\lambda^\downarrow) \, , \qquad 
        \lambda^{*}=\frac{1}{\sqrt{2}}(\lambda^\uparrow-i\lambda^\downarrow) \, ,
   \end{equation*} 
   with
    \begin{equation}
        \lambda^\uparrow(t)=\sum_{n^b \in \mathbb{Z}^3}\lambda^{\uparrow}(t,n^b) \otimes e^{in^b p^b} \,,\qquad
        \lambda^\downarrow(t)=\sum_{r^b \in \mathbb{L}_3}\lambda^\downarrow(t,r^b) \otimes e^{ir^b p^b}\,,
    \label{eq:BFSSMixedBC}
\end{equation}
in terms of the operators $p^a$ introduced in section \ref{sec:ReviewIKKT}, where $\lambda^\uparrow$ and $\lambda^\downarrow$ are real. Here \(\mathbb{L}_3\) is the three-dimensional lattice (\ref{eq:L3}) of half integer modes. When written in this way, all terms involving both components cancel out, and the Lagrangian is decomposed into two non-interacting pieces, one involving periodic spinors and the other anti-periodic
\begin{equation}
    \mathcal{L}_{ferm}=\mathcal{L}^P+\mathcal{L}^{AP} \, .
    \label{eq:FermionicDecomposed}
\end{equation}
Here
\begin{equation*}
    S^P=\int \frac{dt d^3\sigma}{(2\pi L)^{-3}}\mathcal{L}^P=\frac{1}{g^2_{eff}}\int \frac{dt d^3\sigma}{(2\pi L)^{-3}}\Tr\left[\lambda^{\uparrow, T} \mathcal{D}_t \lambda^\uparrow-i\lambda^{\uparrow, T} \sigma^a \mathcal{D}_a \lambda^\uparrow-i\lambda^{\uparrow, T} i\sigma^2\gamma^\mu [X_\mu, \lambda^\uparrow] \right] \, ,
\end{equation*}
and 
\begin{equation*}
     S^{AP}=\int \frac{dt d^3\sigma}{(2\pi L)^{-3}}\mathcal{L}^{AP}=\frac{1}{g^2_{eff}}\int \frac{dt d^3\sigma}{(2\pi L)^{-3}}\Tr\left[\lambda^{\downarrow, T} \mathcal{D}_t \lambda^\downarrow-i\lambda^{\downarrow, T} \sigma^a\mathcal{D}_a \lambda^\downarrow +i\lambda^{\downarrow, T} i\sigma^2\gamma^\mu [X_\mu, \lambda^\downarrow]\right] \, ,
\end{equation*}
where the $\sigma$ dependence comes from the usual mode expansion
\begin{equation*}
    \begin{aligned}
        &A(t,\sigma)= \sum_{n^b \in \mathbb{Z}^3} A(t, n^b) e^{2 \pi iL n^b \sigma^b} \, ,\\
        &X^\mu(t,\sigma)=\sum_{n^b \in \mathbb{Z}^3} X^\mu(t, n^b)e^{2 \pi iL n^b \sigma^b}\,,\\
        &X^a(t, \sigma)=\sum_{n^b \in \mathbb{Z}^3} X^a(t, n^b)e^{2 \pi iL n^b \sigma^b}\,,\\
        &\lambda^\uparrow(t, \sigma)=\sum_{r^b \in \mathbb{Z}^3} \lambda^\uparrow(t, r^b) e^{2 \pi iL r^b \sigma^b}\,,\\
        &\lambda^\downarrow(t, \sigma)=\sum_{r^b \in \mathbb{L}_3} \lambda^\downarrow(t, r^b) e^{2 \pi iL r^b \sigma^b} \, .
    \end{aligned}
\end{equation*}
We plug this back into (\ref{eq:FermionicDecomposed}), Fourier transform with respect to time, and perform the usual rescaling
\begin{equation*}
    X^\alpha \rightarrow k L X^{\alpha}, \quad \lambda^b \rightarrow k L^{3/2} \lambda^b, \quad c \rightarrow k L c,\quad  t\rightarrow \frac{t}{L}, \quad k^2 =\frac{g^2_{eff}}{L^3}\,,
\end{equation*}
with $\alpha=0,...,9$, $b=\uparrow, \downarrow$. It is then straightforward to obtain fermionic propagators from the kinetic terms, which take the form
\begin{equation*}
    \begin{aligned}
        &\left \langle\lambda^\uparrow_{A,\alpha}(p, r^a) \lambda^\uparrow_{B,\beta}(q,s^a)\right \rangle = -i\frac{\delta(p-q)\delta_{r^a,s^a}\delta_{\alpha \beta}(\delta_{AB}\,p+i\sigma^a_{AB}r_a)}{p^2+r_a r^a} \, , \\
        &\left \langle\lambda^\downarrow_{A,\alpha}(p, r^a) \lambda^\downarrow_{B,\beta}(q,s^a)\right \rangle = -i\frac{\delta(p-q)\delta_{r^a,s^a}\delta_{\alpha \beta}(\delta_{AB}\,p+i\sigma^a_{AB}r_a)}{p^2+r_a r^a} \,.
    \end{aligned}
\end{equation*}
We integrate these out to compute expectation values of non-zero modes. The discussion for bosons is unchanged from section \ref{sec:BFSSComp}. On the other hand, integrating out the fermions yields important physical consequences. The effective theory reads, in Lorentzian signature,
\begin{equation}
    S_0^{eff}=S_0-\ln \left \langle e^{-S_{int}}\right \rangle=S_0+\frac{1}{2}\int dt \left[M^2_{X^\mu}\Tr[X_\mu (t,0)X^\mu(t,0)]+M^2_{X^a}\Tr[X_a(t,0)X^a(t,0)]\right] \, ,
    \label{eq:BFSS_Mixed_eff}
\end{equation}
where 
\begin{equation}
    M^2_{X^\mu}= 8\frac{N}{L^2}(S_{F_1}-S_{B_1}), \qquad M^2_{X^a}=\frac{16}{3} \frac{N}{L^2}(S_{F_1}-S_{B_1}) \, ,
    \label{eq:BFSS_Twist_mass}
\end{equation}
and 
\begin{multline*}
     S_0=\frac{1}{g^2_{eff}}\int dt \Tr\left[\frac{1}{2}\mathcal{D}_tX_I(t,0)\mathcal{D}_tX^I(t,0)+\frac{1}{4}[X^I(t,0), X^J(t,0)]^2+i \lambda^{\uparrow, T}(t,0) \mathcal{D}_t\lambda^\uparrow(t,0)\right. \\ \left.
     -\lambda^{\uparrow, T}(t,0) \sigma^a  [X_a(t,0), \lambda^\uparrow(t,0)]-i\lambda^{\uparrow, T}(t,0)i\sigma^2 \gamma^\mu [X_\mu(t,0), \lambda^\uparrow(t,0)]+i \partial_t\bar{c}(t,0) \mathcal{D}_tc(t,0)\vphantom{\frac{1}{2}}\right] \,.
\end{multline*}
We see that periodic fermions cancel with half of the bosonic degrees of freedom in loops and have zero modes, which appear in the effective theory. The anti-periodic fermions do not have zero modes but simply cancel the divergent part of bosonic propagators, thus generating a mass in the effective theory. As a result, we have obtained an effective field theory in which fermions appear explicitly. Bosons have symmetry-breaking masses which are exactly half the masses of the anti-periodic case.

In the next section, we will leverage this construction to study black hole solutions in the theory we have obtained. 
\section{Black hole in the zero-modes theory}
\label{sec:BH}
In this section, we consider possible solutions of the theory (\ref{eq:BFSS_Mixed_eff}) with the correct features to represent a black hole. In particular, we show that the construction of \cite{Chu2024} can be adapted to our effective theory. There, the approach is bottom-up, and the matrix theory deployed is constructed ad hoc to reproduce black hole solutions. Here, the compactification procedure of BFSS with mixed boundary conditions provides a top-down approach, where the theory we study is derived directly from BFSS theory. 
\subsection{Fuzzy sphere background}
The key idea is to consider classical solutions to the bosonic part of (\ref{eq:BFSS_Mixed_eff}) as the background geometry. We then treat fermionic excitations as quantum fluctuations on top of the geometry. We focus on solutions such that
\begin{equation*}
    A=0, \quad c=0, \quad X^\mu=0 \, ,
\end{equation*}
which is made possible by the symmetry-breaking pattern $SO(9) \rightarrow SO(3) \times SO(6)$ we found in section \ref{sec:BFSS_Mixed}. 
Furthermore, we consider static solutions with
\begin{equation*}
    \partial_tX^I=0 \, .
\end{equation*}
From a target space perspective, this amounts to saying that, for those solutions, six spatial coordinates have no extent\footnote{This is true for the semiclassical background, whereas quantum fluctuations are Planck-sized.}, whereas the remaining three transverse directions are macroscopic. Note that the zero-mode effective action obtained as a consequence of torus compactification of the original matrix theory lives in a non-compact target space. 
We can write the relevant part of the Hamiltonian obtained from (\ref{eq:BFSS_Mixed_eff}) as 
\begin{equation*}
	\label{eq:HamBH}
H=R\,k^4\, \Tr\left[\frac{1}{2}P^iP^i-\frac{1}{4}[X^i,X^j]^2-X^i X^i+\lambda^{\uparrow, T}\sigma^i[X_i,\lambda^\uparrow]\right] \, ,
\end{equation*}
where we have changed units to highlight the dependence on the light-like compactification radius $R$ in M-theory (see \cite{Banks_1996}) and rescaled the fields according to their scaling dimension
\begin{equation}
X^i \rightarrow k X^i, \quad \lambda^\uparrow(t,0) \rightarrow k^{\frac{3}{2}} \lambda^\uparrow(t,0),\quad P^i \rightarrow k^2P^i,\quad k^2=\frac{\sqrt{R}\,M^2_{X^i}}{2} \, .
\label{eq:kdef}
\end{equation}
The classical equations of motion for the bosonic sector, for a static solution, read
\begin{equation}
    \left[X^i,[X^i, X^j]\right]=2 X^j \, , \quad P^i=\partial_tX^i=0 \, ,
    \label{eq:comm_identity}
\end{equation}
where $i=1,2,3$. These equations are solved by the $s=\frac{N-1}{2}$ representation of $SU(2)$, which defines a fuzzy sphere through
\begin{equation}
    [X^i, X^j]=i \epsilon^{ijk} X^k \, ,
    \label{eq:Fuzzysph}
\end{equation}
along with the Casimir relation 
\begin{equation*}
    \sum_i X^{i \, 2}= \frac{N^2-1}{4} \mathbb{I}_N \,.
\end{equation*}
As we learn from \cite{Banks:1997, Banks:1997hz, Horowitz:1997}, any black hole solution in matrix theory needs to fit into the longitudinal direction. This requires a boost, which expands the longitudinal dimension at the horizon to fit the black hole, such that the momentum in the longitudinal direction is non-zero. This momentum is quantised in terms of the light-like radius $R$
\[
P^+=\frac{N}{R} \, .
\]
The net effect is a solution living in five total dimensions, one of which is longitudinal and has size $R$ before the boost. As explained above, the symmetry-breaking pattern \(SO(9) \rightarrow SO(3) \times SO(6)\) allows for solutions which only extend in three space directions. We consider the other space directions to have a Planck-sized extent. In the target space description, they are integrated over and produce a volume factor \(V_{6d}=l_p^6\), $l_p$ being the Planck length in $11$-dimensions. As a consequence, starting from the eleven-dimensional Newton constant \(G_{11}=l_p^{9}\), we get the five-dimensional one through 
\[
G_5=\frac{G_{11}}{V_{6d}}=l_p^3\,.
\] 
Let us now consider the fermionic term
\begin{equation}
    \mathcal{H}_{ferm}= R\,k^4 \Tr\left[\lambda^{\uparrow, \dagger}_{\alpha, a}(t,0)\sigma^j_{ab} [X_j(t,0), \lambda^{\uparrow, \alpha}_b(t,0)]\right],
    \label{eq:fermionicterm}
\end{equation}
where $\lambda^{\uparrow, \dagger}=\lambda^{\uparrow, T}$ because it is a real spinor, and from now on we drop the $\uparrow$ superscript for convenience. We recall from section \ref{sec:BFSS_Mixed} that spinors carry indices \(a=1,2\) and \(\alpha=1,...,4\), which we have made explicit for clarity. On the fuzzy sphere background, this can be arranged as a collection of fermion harmonic oscillators, given by 
\begin{equation}
\mathcal{H}_{ferm}=R\,k^4\sum_{n=1}^{N-1}\left(\sum_{r=1}^{2n+2}\sum_{\alpha=1}^4\lambda_{+,n} \xi^{r,n, \alpha \dagger} \xi^{r,n}_\alpha+\sum_{r=1}^{2n}\sum_{\alpha=1}^4|\lambda_{-, n}| \chi^{r,n, \alpha \dagger} \chi^{r,n}_\alpha-8n(n+1) \right)\, .
\label{eq:Fermenergy}
\end{equation}
Here \(\lambda_{+, n}, \, \, \lambda_{-,n}\) are the eigenvalues of the operator 
\[
\mathrm{X}:=\sigma^j[X_j, \cdot] \, ,
\]
given by
\[
\lambda_{\pm,n}=-\frac{1}{2}\pm\left(n+\frac{1}{2}\right) \, ,
\]
and \(\xi^{r,n}_\alpha, \, \, \chi^{r,n}_\alpha\) are fermionic annihilation operators, obtained by projecting the matrices \(\lambda^\uparrow\) on the corresponding eigenvectors.
We prove this result in subsection \ref{sec:fermFuzzy}.

We recognise (\ref{eq:Fermenergy}) as the Hamiltonian of a set of fermionic oscillators with increasing frequencies, given by $\omega^{\pm}_n=R\,k^4|\lambda_{\pm, n}|$. Since we are working at zero temperature, a state in this system is constructed by progressively filling up the Fermi sea, starting from the lowest energy shell and occupying shells of increasing energies as we increase the number of excited oscillators. A state of given energy is thus constructed by filling up some energy shells completely and partially filling up the highest contributing energy shell. The degeneracy $\Omega$ of such a state is then given by the number of inequivalent options there are to partially fill this shell. In principle, we should also include bosonic quantum fluctuations. We will analyse them in section \ref{sec:bosonicfluc} and show that they do not contribute to our construction.
\subsection{Black Hole solution}
\label{sec:BHsol}
In order to have a black hole solution, we must fill the shell of energy $|\lambda_{-, N-2}|$, which excites $j=y(16N-16)$ oscillators, $0<y<1$. We provide more details in Appendix \ref{sec:fermisea}.

In the boosted frame, where the system has a non-trivial $P^+$, the Hamiltonian gives the lightcone energy. This is related to the invariant mass through 
\[
H=E_{LC}=\frac{R}{2N}M^2_{5D} \,,
\]
where $M_{5D}$ is the mass of the Schwarzschild black hole in five dimensions\footnote{Recall that we are working in units such that the Planck length is one.}, which is related to the Schwarzschild radius, $r_0$, by
\[
M_{5D}=r_0^2 \, .
\]
The bosonic part of the Hamiltonian yields, when evaluated on the fuzzy sphere background,
\[
H_{bos}=-\frac{R\, k^2}{2}NR^2_f \, ,
\]
where $R_f=\frac{k\,N}{2}$ is the physical fuzzy sphere radius in the large $N$ limit. This can be added to the fermionic energy computed in the Appendix \ref{sec:fermisea} such that the total energy, at leading order in $N$, reads 
\[
E_{LC}=\frac{61R\, k^2}{6}NR^2_f \, ,
\]
and the condition that the fuzzy sphere sits behind its Schwarzschild radius,
\[
r_0\geq R_f \, ,
\]
is satisfied.
Since, given the fuzzy sphere background (\ref{eq:Fuzzysph}), all the degeneracy is captured by the fermionic sector of the theory, we can compute the entropy from such degeneracy. We are free to choose the fraction \(y\) of excited fermions in the highest shell. Let us consider the half-filled case \(y=\frac{1}{2}\). Then
\[
S=\log \Omega=\log{\binom{16N-16}{y(16N-16)}}\approx \gamma (N-1)+O(\log N), \qquad \gamma \approx 11.1 \, ,
\]
scales linearly with $N$ in the large $N$ limit. The coefficient $\gamma$ comes from Stirling's approximation
\[
\log \binom{16N-16}{y(16N-16)}
\approx
16(N-1)H(y)
-\frac12 \log\!\bigl(32\pi (N-1)y(1-y)\bigr)
\]
with 
\[
H(y)=\left[-y\log y-(1-y)\log(1-y)\right] \, .
\]
The appropriate entropy-radius relation for this configuration, namely
\[
S\propto r_0^3 \, ,
\]
is only satisfied if $r_0 \sim N^{\frac{1}{3}}$, which requires $k \propto N^{-\frac{2}{3}}$. Note that the scaling of $k$ follows from the mass term, which is generated through compactification, see equation (\ref{eq:kdef}). This, in turn, is related to the torus dimension $L$ (eq. (\ref{eq:BFSS_Twist_mass})). It follows that the compactification limit and the large $N$ limit should be taken in a double-scaled fashion, such that
\begin{equation}
\label{eq:DoubleScal}
L\propto N^{\frac{7}{6}}, \qquad N \rightarrow \infty \, .
\end{equation}
Note that, under this condition, the logarithmic correction \(S_{log} \propto \log r_0^3\) is also reproduced.

The scaling of $S \propto N$ is the one expected for the transition between a black hole localised in the longitudinal direction and a black string stretched along it \cite{Horowitz:1997}. Furthermore, if the scaling is $S\propto N^{\beta}$, $\beta \uparrow 1$, we are approaching the transition, corresponding to the Gregory–Laflamme instability from the black hole state to the black string. In our model, we can construct solutions with lower entropy by choosing a larger fraction of fermionic oscillators to be excited in the largest shell of the Fermi sea $y>\frac{1}{2}$. This, in turn, causes the Schwarzschild radius to increase compared to the fuzzy sphere radius. 

We can, in particular, realise a black hole configuration with entropy $S \propto N^{\alpha}, \,0<\alpha<1$ by exciting a fraction $y$ of oscillators in the leading shell, such that $y$ tends to $1$ parametrically with $N$. In this case, the degeneracy given by Stirling's formula for large $N$ is
\[
\Omega \sim \frac{e^{16NH(y)}}{\sqrt{2\pi 16N y(1-y)}}, \quad H(y)=-y\log{y}-(1-y)\log{(1-y)}, \quad 0<y<1 \, .
\]
Choosing 
\[
y=1-\frac{cN^{\alpha-1}}{\log{N}}, \quad 0<\alpha<1 \, ,
\]
$c$ being an arbitrary constant of order $1$, we obtain 
\[
S=\log{\Omega}=16NH(y)-\frac12 \log\!\bigl(32\pi Ny(1-y)\bigr), \quad H(y) \sim (1-\alpha)c N^{\alpha-1} \, ,
\]
at leading order in $N$. Again, applying the entropy-radius relation and the relation between the Schwarzschild radius and the generated mass, we obtain the corresponding scaling of $k$ as
\[
k\propto N^{\frac{\alpha}{3}-1} \Rightarrow L \propto N^{\frac{3}{2}-\frac{\alpha}{3}} \, .
\]
Our approach, therefore, shows that it is possible to construct a black hole solution as a fuzzy sphere background plus fermions in a state that generates the black hole entropy. We saw that, by taking the decompactification limit in the original theory in a specific way, see equation (\ref{eq:DoubleScal}), we can approach the limiting case of the black hole/black string transition, represented by the entropy scaling like $S \sim N$. Moreover, we recover the known relations between physical quantities characterising a black hole.

The fuzzy sphere solution we found in this section is still valid even for different values of $L$ and, therefore, of the mass parameter $M_X$, which do not saturate the Bekenstein bound on the entropy. These solutions cannot consistently represent black holes, which can be seen by looking at the boost along the compact dimension. We can show that the condition on the entropy to be smaller than the Bekenstein bound puts a constraint on the boost parameter. Consider the case where the entropy is maximum \(S \sim\, N\), we have 
\[
P^+=\frac{N}{R}=\frac{S}{R} \, .
\]
Such momentum is obtained through a longitudinal boost, in terms of which it reads \cite{Horowitz:1997}
\[
P^+=r_0^2\, e^{\rho} \, ,
\]
where \(\rho\) is the boost parameter. If the entropy does not saturate the Bekenstein bound, we have 
\[
S < r_0^3 \Rightarrow e^\rho <\frac{r_0}{R} \, ,
\]
for a large boost. This implies that the longitudinal direction after the boost is smaller than the Schwarzschild radius, and the localised black hole interpretation breaks down. More generally, in the case of \(S \propto N^\alpha\), configurations that give \(S<r_0^3\) correspond to solutions which are not consistent with the black hole interpretation. Therefore, we can only interpret our fuzzy sphere plus fermions construction as a black hole if the torus radius scales with \(N\) appropriately and the Bekenstein bound is saturated.

\subsection{Fermionic theory on the fuzzy sphere background}
\label{sec:fermFuzzy}
In this section, we prove that the fermionic Hamiltonian consists of a collection of harmonic oscillators when evaluated on the fuzzy sphere background. We consider the operator 
\begin{equation*}
    \mathrm{X}:=\sigma_i[X^i, \cdot], \quad \mathrm{X}: \mathbb{C}^2 \times End(\mathcal{V}_s) \rightarrow \mathbb{C}^2 \times End(\mathcal{V}_s) \, ,
\end{equation*}
where the commutator \([X^i, \cdot]\) acts on the space of traceless $N \times N$ matrices as endomorphisms of the spin $s=\frac{N-1}{2}$ representation \(\mathcal{V}_s\). The $\mathbb{C}^2$ factor is the space of two-dimensional complex spinors acted on by $\sigma^i$. Altogether, \(\mathrm{X}\) acts on a \(2(N^2-1) \)-dimensional complex space. We call this an adjoint-like representation, constructed from the spin $s$ representation of $SU(2)$.

Let us now consider the square operator
\begin{equation*}
   \mathrm{X}^2=\sigma_i\sigma_j\left[X^i, [X^j, \cdot]\right] \, ,
\end{equation*}
which, as a consequence of (\ref{eq:Fuzzysph}), satisfies
\begin{equation}
    \mathrm{X}^2+\mathrm{X}-\left[X^i,[X^i, \cdot]\right]=0 \, ,
    \label{eq:Xeq}
\end{equation}
where the operator
\[
\Delta:=\left[X^i,[X^i, \cdot]\right]
\]
is the Casimir operator of the adjoint-like representation and defines a generalisation of the Laplacian to the non-commutative fuzzy sphere geometry. Indeed, this acts as the identity on the $\mathbb{C}^2$ component. We can decompose
\[
End(\mathcal{V}_s) \cong \mathcal{V}_s \otimes \mathcal{V}^*_s \cong \mathcal{V}_s \otimes \mathcal{V}_s \, ,
\]
where the second isomorphism comes from the fact that $SU(2)$ representations are real. It follows that we can carry out a Clebsch-Gordan decomposition 
\begin{equation}
End(\mathcal{V}_s) \cong \bigoplus_{k=0}^{2s}\mathcal{V}_k \, ,
\label{eq:CG}
\end{equation}
such that the Casimir operator acts on each factor as
\begin{equation}
\Delta|_{\mathcal{V}_k}=k(k+1) \mathbb{I}_{2k+1} \, ,
\label{eq:Casimir}
\end{equation}
where the dimension of each Clebsch-Gordan factor is $dim(\mathcal{V}_k)=2k+1$. It follows that, given an eigenvector $u_{\lambda}$ of $\mathrm{X}$, equation (\ref{eq:Xeq}) reads
\begin{equation*}
     \lambda^2 u_\lambda+\lambda u_\lambda-\Delta u_\lambda=0 \, .
\end{equation*}
We decompose $u_\lambda$ consistently with (\ref{eq:CG}) 
\[
u_\lambda=
v_1+...+v_{2s} \, ,
\]
and use (\ref{eq:Casimir}) (note that $v_0$ is not present because of the tracelessness condition). As a consequence of linear independence, we get
\[
\left(\lambda^2+\lambda-k(k+1)\right)v_k=0, \quad k=1, \dots, 2s\,.
\] 
We see that, for each $k$, $u_\lambda$ can be decomposed into $v_{k\pm} \in \mathcal{V}_k$, lying respectively in spin $k\pm\frac{1}{2}$  representations of $SU(2)$, and having eigenvalues\footnote{Note that the total dimension of this space is indeed
\[\sum_{k=1}^{N-1} 4k+2=2(N^2-1)\,,\]
and that 
\[
\Tr(\mathrm X|_{\mathcal{V}_k})=
(2k+2)k+2k(-k-1)=0\,,
\]
as is required by the traceless structure of $\mathrm{X}$. }
\[
\lambda_{\pm,k}=-\frac{1}{2}\pm\left(k+\frac{1}{2}\right) \, .
\]
It follows that the spectral decomposition of $\mathrm{X}|_{\mathcal{V}_n}$ in terms of eigenvalues and eigenvectors is given by
\[
\mathrm{X}|_{\mathcal{V}_n}=\sum_{r=1}^{2n+2}\lambda_{+,n}U^{r,n\dagger}_{a \alpha \gamma}U^{r,n}_{b \beta \sigma}+\sum_{r=1}^{2n}\lambda_{-,n}V^{r,n \dagger}_{a \alpha \gamma}V^{r,n}_{b \beta \sigma} \, ,
\]
where $U^{r,n}$ and $V^{r,n}$ are eigenvectors corresponding to positive and negative eigenvalues respectively, obtained by representing $\mathcal{V}_n$ in the space of $N\times N$ matrices, such that $a, b=1,2$ are spinor indices and $\alpha, \beta, \gamma, \sigma=1,..., N$ are matrix indices. From (\ref{eq:CG}), the full operator is then
\[
\mathrm{X}=\sum_{\substack{n=1}}^{N-1}\mathrm{X}|_{\mathcal{V}_n}=\sum_{\substack{n=1 }}^{N-1}\left(\sum_{r=1}^{2n+2}\lambda_{+,n}U^{r,n\dagger}_{a \alpha \gamma}U^{r,n}_{b \beta \sigma}+\sum_{r=1}^{2n}\lambda_{-,n}V^{r,n \dagger}_{a \alpha \gamma}V^{r,n}_{b \beta \sigma} \right)\, .
\]
Having obtained a spectral decomposition for the operator, we substitute it back into the fermionic Lagrangian (\ref{eq:fermionicterm}). This allows us to define fermionic oscillators
\[
\xi^{r,n}_\alpha := U^{r,n}_{b\beta \sigma} \lambda^{\uparrow}_{b\beta \sigma, \alpha}, \quad \chi^{r,n \dagger}_\alpha:=V^{r,n}_{b\beta\sigma} \lambda^\uparrow_{b\beta\sigma, \alpha}\,,
\]
where the indices take respective values \(b=1,2\) for \(SO(3)\) spinor indices, \(\beta, \sigma=1,\dots,N\) for matrix \(SU(N)\) indices, and \(\alpha=1, \dots, 4\) for \(SO(6)\) spinor indices. These oscillators satisfy canonical anti-commutation relations
\[
\{\xi^{r,n}_\alpha, \xi^{q,m, \gamma \dagger}\}=\delta^{rq}\delta^{n,m}\delta_{\alpha}^\gamma, \quad \{\chi^{r,n}_\alpha, \chi^{q,m,\gamma \dagger}\}=\delta^{rq}\delta^{nm}\delta_\alpha^\gamma.
\]
As a consequence, the fermionic Hamiltonian reads
\begin{equation*}
\mathcal{H}_{ferm}=R\,k^4\sum_{n=1}^{N-1}\left(\sum_{r=1}^{2n+2}\sum_{\alpha=1}^4\lambda_{+,n} \xi^{r,n, \alpha \dagger} \xi^{r,n}_\alpha+\sum_{r=1}^{2n}\sum_{\alpha=1}^4|\lambda_{-, n}| \chi^{r,n, \alpha \dagger} \chi^{r,n}_\alpha-8n(n+1) \right)\,,
\end{equation*}
which proves the claim.
\subsection{Bosonic fluctuations}
\label{sec:bosonicfluc}
In this section, we show that bosonic fluctuations do not affect our black hole construction. First, we analyse semiclassical fluctuations of the background geometry. We will show that the solution is stable under those perturbations. Then, we consider quantum bosonic fluctuations and show that they do not contribute significantly to the black hole solution, in the large \(N\) limit.
\subsubsection{Classical bosonic fluctuations}
The analysis of semiclassical fluctuations of the fuzzy sphere can be adapted from \cite{Chu2024}. The idea is to consider \(X^i+\delta X^i\), where \(X^i\) define the fuzzy sphere geometry (\ref{eq:Fuzzysph}) and \(\delta X^i\) are perturbations. Then, at quadratic order in the perturbations, the Lagrangian for fluctuations can be read off from (\ref{eq:HamBH})
\[
\mathcal{L}_{\delta X^i}=R\,k^4\, \Tr\left[\frac{1}{2}\delta \dot{X}^i\delta \dot{X}^i+\frac{1}{4}\left([X^i, \delta X^j]-[X^j, \delta X^i]\right)^2+\frac{1}{2}[X^i,X^j][\delta X^i, \delta X^j]+\delta X^i \delta X^i\right] .
\]
We can define the derivative operator 
\[
L^i := [X^i, \cdot] \, ,
\]
which, following the argument of section \ref{sec:fermFuzzy}, is an angular momentum operator. It therefore has an eigenbasis  \( | \ell, m_z \rangle\) with
\[
L^2 | \ell, m_z \rangle = \ell (\ell +1) | \ell, m_z \rangle, \quad L_3| \ell, m_z \rangle=m_z | \ell, m_z \rangle\,,\quad
0\leq \ell \leq N-1, \quad -\ell \leq m_z \leq \ell\,.
\]
The angular momentum \(\ell\) is truncated due to the fuzziness of the sphere. Moreover, we can define \((\varepsilon_i)_{jk} := -i \epsilon_{ijk}\), which is an angular momentum with \(\ell_\varepsilon=1\), since 
\[
[\varepsilon_i, \varepsilon_j]=i\epsilon_{ijk} \, \varepsilon_k, \quad \varepsilon^i \varepsilon^i=2 \, .
\]
As shown in \cite{Chu2024}, the Lagrangian in terms of these operators reads
\begin{equation}
\mathcal{L}_{\delta X^i}=R\,k^4\, \Tr\left[\frac{1}{2}\delta \dot{X}^i\delta \dot{X}^i-\frac{1}{2}\delta X^i N_{ij} \delta X^j\right] \, ,
\label{eq:fluctuationparallel}
\end{equation}
where 
\[
N_{ij}=\left((\varepsilon \cdot L)^2-(\varepsilon \cdot L)-2\right).
\]
It follows from the equation of motion 
\[
\delta \Ddot{X}_i+N_{ij} \delta X_j=0 
\]
that the problem of the stability of the various fluctuation modes becomes the analysis of the eigenvalues of $N_{ij}$. In particular, positive eigenvalues correspond to stable modes, whereas negative eigenvalues signal unstable perturbations. Note that
\[
\varepsilon \cdot L=\frac{1}{2}\left(J(J+1)-\ell(\ell+1)-2\right) \, ,
\]
where $J$ corresponds to the total angular momentum \(J^i=L^i+\varepsilon^i\). Then the eigenvalue problem 
\[
N_{ij}\phi^j=\lambda \phi^i
\]
yields eigenvalues 
\begin{equation}
    \lambda=\begin{cases}
        -2 & \quad \text{for} \, \, \ell=0, J=1\text{~and~}\ell=1, J=2 \\
         0 & \quad \text{for} \, \, \ell \geq 1, J=\ell \text{~and~} \ell=2, J=3\\
        \ell(\ell+3) & \quad \text{for} \, \, \ell \geq 1, J=\ell-1 \\
        (\ell+1)(\ell-2) & \quad \text{for} \, \, \ell \geq 3, J=\ell+1
    \end{cases} 
    \label{eq:modesparallel}
\end{equation}
and eigenfunctions given by the Clebsch-Gordan composition of the polarisation vector \(\varepsilon^p\) and spherical harmonics \(\hat{Y}^\ell_{m_z}\)
\[
\phi^i=\sum_{m_z=-\ell}^\ell\sum_{p=0,\pm1}C^{J, M_Z}_{\ell,m_z;1,p} \varepsilon^p_i\hat{Y}^\ell_{m_z} \, .
\]
Modes with $ \ell \geq 1, J=\ell-1 $ and $\ell \geq 3, J=\ell+1$ have positive eigenvalues, and therefore represent stable fluctuations. Two modes have negative eigenvalues; the mode
 \(\ell=0, J=1\), corresponds to an overall \(U(1)\) shift, is absent because the \(X^i\) are traceless; \(\ell=1, J=2\), describes fluctuations with
 \[
 \delta X^i = S^i_j X^j
 \] 
 called squashing modes, where \(S^i_j\) is a symmetric traceless tensor and \(X^i\) are the fuzzy sphere background matrices. We show in Appendix \ref{sec:stability} that the squashing modes are rendered stable by the corresponding fluctuations in the fermionic sector. In particular, we compute the potential, quadratic in the fluctuation, generated by the perturbation in the fermionic sector, in the state given by the Fermi sea with partially filled highest shell. We show that the sum of the bosonic and fermionic contributions gives a positive potential, corresponding to a stable perturbation.
 
Regarding zero-modes, those with \(J=\ell\) represent gauge transformations, which are not physical perturbations. This can be seen by first counting the number of such modes 
\[
\#_{zero}= \sum_{\ell=1}^{N-1}(2\ell+1) = N^2 -1 \,,
\]
which matches the number of independent generators. Moreover, an infinitesimal gauge transformation takes the form 
\[
\delta X^i =i[X^i, \Lambda] = i L^i\Lambda  \, ,
\]
which is an eigenfunction of \(\epsilon \cdot L\) with eigenvalue \(-1\). We can check this explicitly
\[
(\epsilon \cdot L)_{kl} L^l \Lambda =-i \epsilon_{jkl} L^j L^l=\frac{1}{2}\epsilon_{jkl} \epsilon_{jl m} L^m = -L^k\,,
\]
which proves the statement. The remaining zero-modes, with \(\ell=2, J=3\), represent marginal fluctuations. However, the quartic potential can be evaluated, and it provides a positive contribution, which makes the perturbations marginally stable.

A similar analysis can be done for fluctuations in the directions orthogonal to the fuzzy sphere, to check that they also remain stable. The background geometry in this case is given by \(X^a=0, \,a=4,...,9\), and we are interested in perturbations \(\delta X^a\). The perturbation Lagrangian at quadratic order takes a much simpler form 
\begin{equation*}
\mathcal{L}_{\delta X^a}=Rk^4 \Tr \left[\frac{1}{2}\delta \dot{X}^a \delta \dot{X}^a+\frac{1}{2}[X^i, \delta X^a][X^i, \delta X^a]+\frac{3}{2}\delta X^a \delta X^a\right] \, .
\end{equation*}
The potential can be further simplified by noting that
\[
\Tr \left[[X^i, \delta X^a][X^i, \delta X^a]\right]=-\Tr\left[\delta X^a \bigl[X^i, [X^i, \delta X^a]\bigr]\right]=-\Tr\left[\delta X^a L^2 \delta X^a\right] \, ,
\]
which results in 
\begin{equation}
\label{eq:fluctuationtransverse}
    \mathcal{L}_{\delta X^a}=Rk^4 \Tr \left[\frac{1}{2}\delta \dot{X}^a \delta \dot{X}^a-\frac{1}{2}\delta X^a(L^2-3)\delta X^a\right]\,.
\end{equation}
As a consequence, the equations of motion have the simple form 
\[
\delta \ddot{X}^a+\left(L^2-3\right) \delta X^a=0\,,
\]
and the problem of the stability of fluctuations reduces to the positivity of the eigenvalues of the operator \(\left(L^2-3\right)\). The eigenvalue problem 
\begin{equation}
\label{eq:bos_fluc_ort}
\left(L^2-3\right) \phi =\mu \,\phi
\end{equation}
is solved by the spherical harmonics with eigenvalues determined by quantum numbers \(0\leq \ell\leq N-1, \, -\ell\leq m_z \leq \ell\), such that 
\begin{equation}
    \phi=\hat{Y}^\ell_{m_z}, \quad \mu= \ell(\ell+1)-3 \, .
    \label{eq:modestransverse}
\end{equation}
This implies that almost all the eigenvalues are positive, and therefore correspond to stable perturbations. The negative eigenvalues are: the \(\ell=0\) mode, which represents a \(U(1)\) shift, which is forbidden for traceless matrices; the \(\ell =1\) modes, which, similarly to the case of \(\ell=1, J=2\) fluctuations above, corresponds to squashing modes \[
\delta X^{i+3}= A^i_j X^j \, ,
\]
where \(A^i_j\) is the \(3 \times 3\), anti-symmetric matrix parametrising the fluctuation. The instability is fixed by including perturbations in the fermionic sector, analogously to the case above for perturbations along the fuzzy sphere directions, as shown in Appendix \ref{sec:stability}. Note that the stability of fluctuations strongly depends on the ratio between the two types of masses (\ref{eq:BFSS_Twist_mass}) associated with bosonic matrices, which is a consequence of the top-down derivation from BFSS through torus compactification. In fact, the \(-3\) contribution appearing in (\ref{eq:bos_fluc_ort}) depends on that ratio directly. More negative values would lead to unstable fluctuations.
The fuzzy sphere background is, therefore, stable against classical bosonic fluctuations.
\subsubsection{Quantum bosonic fluctuations}
The quadratic potentials in (\ref{eq:fluctuationparallel}) and (\ref{eq:fluctuationtransverse}) for bosonic fluctuations can be quantised as simple harmonic oscillators. Since those terms represent fluctuations on top of the fuzzy sphere background, we subtract the zero-point energy from the quantised Hamiltonian, such that the energy of the bosonic ground state is that of the fuzzy sphere. We decompose the fluctuations \(\delta X^i, \,\delta X^a\) into eigenmodes of the respective quadratic potential, and quantise every eigenmode independently. The quantum Hamiltonian for the fluctuations reads
\[
H_{\delta X}=Rk^4 \sum_{\ell=1}^{N-1}\sum_{J=\ell-1, \ell+1}\sum_{m_z=-J}^J\sqrt{\lambda_{\ell, J}} \,a^\dagger_{\ell, J,m_z} a_{\ell, J,m_z}+R k^4 \sum_{\ell=1}^{N-1}\sum_{m_z=-\ell}^\ell\sqrt{\mu_{\ell}} \, b^\dagger_{\ell, m_z} b_{\ell, m_z} \, .
\]
Here, \(\lambda_{\ell, J}\) and \(\mu_{\ell}\) are the eigenvalues of the classical bosonic modes (\ref{eq:modesparallel}), (\ref{eq:modestransverse}) and \(a_{\ell, J,m_z}\), \(b_{\ell, m_z}\) are the annihilation operators for each mode. The theory is at zero temperature, which implies that bosonic fluctuations are frozen in the ground state and do not contribute to the energy and entropy of the black hole solution we constructed. We conclude that bosonic fluctuations do not affect our solution, which is composed of a fuzzy sphere background with fermionic excitations which partially fill up the Fermi sea, and bosonic harmonics in the overall ground state. It would be interesting to study the effect of thermal fluctuations in the case where \(T \neq 0\). We leave this to future work.
\section{Conclusion}
\label{sec:Conclusion}
In this paper, we analysed torus compactifications of IKKT and BFSS theories as a mechanism for the generation of a mass term. Building on recent results, we imposed a novel set of boundary conditions on fermionic degrees of freedom in IKKT theory. These amount to acting with a non-trivial operator on the spinor when winding around any of the torus dimensions. In the case of IKKT, we analysed two kinds of mixed boundary conditions. In both cases, the action on the spinor is not a symmetry of the compact theory. As a consequence, some fermionic degrees of freedom decouple and do not contribute to loop diagrams, thus failing to cancel bosonic divergences. Hence, the theory requires some \emph{ad hoc} renormalisation. In the decompactification limit, non-zero modes become heavy and can be integrated out, thus obtaining an effective theory for zero modes. Such a theory breaks the original $SO(1,9)$ symmetry to a subgroup of $SO(1,3)\times SO(6)$, where the $SO(1,3)$ itself is broken to a subgroup.

We then analysed BFSS theory, firstly with anti-periodic boundary conditions, and showed that the resulting effective theory generates a mass term and all fermions are quenched. Then, we considered mixed boundary conditions. In this case, we were able to act on spinors with a symmetry operator when winding around the torus. As a consequence, divergences cancel automatically, and no renormalisation is required. The result is an effective theory with broken $SO(9) \rightarrow SO(3) \times SO(6)$, a mass term, and half the fermionic degrees of freedom compared to the initial theory.

In the last part of the paper, we took this top-down construction as a starting point for explicitly characterising black hole solutions using matrix theory degrees of freedom. We found a configuration where a black hole spacetime is described by a fuzzy two-sphere, while fermionic excitations and their degeneracies give rise to the black hole entropy. Moreover, we showed how classical bosonic perturbations are stable and do not disrupt the fuzzy sphere configuration. Their quantum counterparts are frozen in the ground state and do not contribute to the solution. 

This construction provides a direct realisation of a black hole in matrix theory, with features that are consistent with previous studies of Schwarzschild black holes in the M-theory region of BFSS theory. In particular, we reproduce the $S \leq c \, N$ scaling of the entropy, \(c=O(1)\) being a constant. The entropy-to-radius relation is also reproduced, including logarithmic corrections, but this requires the torus dimension $L$ to go to infinity as a power of the matrix size $N$, in the decompactification limit. 

It would be interesting to investigate the saturated value of the entropy further, in order to assess whether the black hole/black string transition can be captured by this construction. In particular, a possible future direction would be the explicit construction of black string solutions from matrix degrees of freedom.

Fuzzy sphere solutions have been considered in IKKT in connection to the emergence of gravity from the matrix model, see e.g. \cite{Steinacker:2016, Komatsu:2024, Hartnoll:2024}. As an interesting future direction, one could try to replicate the analysis we carried out in the context of BFSS theory to the zero-mode effective theory obtained through torus compactification with mixed fermionic boundary conditions of IKKT. The main difference to the polarised IKKT model \cite{Hartnoll:2024} is that the mass deformation derived through compactification breaks supersymmetry of the compact action. It would be interesting to explore the consequences of this.

In this work, we have considered BFSS theory at zero temperature, in the M-theory region of parameter space. It would be interesting to repeat the analysis for generic temperature and to look at other regions of parameter space. For instance, one could consider the 't Hooft limit of BFSS theory at higher temperature, where it is known to be dual to the black brane solution of type IIA supergravity \cite{Itzhaki_1998}. The evaporation of the black brane has been fully characterised in matrix theory \cite{Berkowitz_2016} and shown to reproduce a Page curve for the entanglement entropy \cite{Choudhury:2024hoh}. It would be interesting to extend the construction presented in this paper to the supergravity regime of the parameter space. Achieving such a result would open the possibility of studying the black hole evaporation mechanism explicitly. This, in turn, would provide full theoretical control over the time-dependent dynamics of the process, offering valuable insight into the information paradox. 
\acknowledgments 
DL acknowledges funding from the STFC studentship grant ST/X508664/1. This work was presented as a poster at Eurostrings 2026 in Athens, Greece, and we thank the conference participants for stimulating discussions. DL thanks the theoretical physics group at the Institute of Physics of the Czech Academy of Sciences for the invitation to present this work, and acknowledges useful discussions with Paolo Rossi, Andrea de Marco, and Prof. Joris Raeymaekers. We also thank Lucas Leung and Christopher Butcher for valuable discussions.
\appendix
\section{Alternative block decomposition}
\label{sec:torusdecomposition}
We reformulate the block decomposition of matrix degrees of freedom introduced in section \ref{sec:ReviewIKKT} and used throughout the paper. In particular, we will adopt an explicit matrix decomposition which highlights the block matrix structure of bosonic terms and realises the mode decomposition for anti-periodic fermions. 

We have seen how compactifying Matrix Theories on a torus amounts to replicating the fundamental region an infinite number of times. This is implemented by splitting the Hilbert space of matrix degrees of freedom into two components
\[
X=Y \otimes Z \, ,
\]
where \(Y\) represents the degrees of freedom in the fundamental region and \(Z\) contains information about transitions between different regions. Consider the decomposition of the matrix 
\[
A^\mu = \sum_{n^b \in \mathbb{Z}^6} A^\mu (n^b) \otimes e^{in^b p^b} \, .
\]
We want to provide an explicit matrix representation for the operator \(e^{in^b p^b}\). Let us take the one-dimensional example for simplicity of notation. The higher-dimensional case follows immediately, since every torus direction is parametrised independently. The natural choice is the following:
\[
\left(e^{in p} \right)_{ij}:= C^n_{ij}= \delta_{i, j+n} \, ,
\]
which represents a matrix with entries \(1\) on the diagonal at distance \(n\) from the central one and \(0\) elsewhere. Moreover, it satisfies
\[
C^n C^m =C^{n+m} \, . 
\]
Such matrices give an explicit representation of operators \(e^{inp}\) in the \( | q\rangle \) basis of the Hilbert space \(Z\).
The decomposition in winding modes now reads 
\[
A^\mu = \sum_{n \in \mathbb{Z}} A^\mu (n) \otimes C^n \, .
\]
Similarly, we define a matrix \(Q\) which labels the fundamental regions 
\[
Q_{ij}= j \,\delta_{ij}
\]
and exponentiates to 
\[
\left(e^{-2\pi i Q}\right)_{kj}=e^{-2\pi i j} \delta_{kj}= \delta_{kj} \, ,
\]
where the last equality simply follows from the fact that matrix indices are integers. Now, the matrix corresponding to compact direction is decomposed as 
\[
A^{9} = \sum_{n \in \mathbb{Z}} A^{9} (n) \otimes C^n+2\pi L \mathbb{I}_N Q\,.
\]
The commutator with $C^n$ is
\begin{equation}
\label{eq:commutation}
    [Q, C^n]= n C^n \, .
\end{equation}
It follows that the translation operator can be written as
\[
U= \mathbb{I}_N \otimes e^{-2\pi i Q}C^1= \mathbb{I}_N \otimes C^1\,,
\] 
then the action on the matrices has the desired form
\[
\begin{aligned}
    (U)^{-1}A^{\mu} U&=A^{\mu} \\
    (U)^{-1}A^{9} U&=A^{9}+2\pi L \mathbb{I}_N\otimes \mathbb{I}_M \, .
\end{aligned}
\]
Moreover, when fermions are periodic, their decomposition is straightforward 
\[
\psi = \sum_{n \in \mathbb{Z}} \psi(n) \otimes C^n \, .
\]

We now extend the construction to accommodate anti-periodic fermions. This requires doubling the Hilbert space, such that 
\[
Z= \mathcal{H}_{\mathbb{Z}}\oplus\mathcal{H}_{\mathbb{Z}+\frac{1}{2}} \, .
\]
Namely, we double the winding modes lattice to account for both integer and half-integer modes, which take value in the respective lattices. Anti-periodic fermions are given by the block matrix 
\[
\psi =
\begin{pmatrix}
0 & \psi(-\frac12) & 0 & \psi(-\frac32) & 0 & \cdots \\
\psi(\frac12) & 0 & \psi(-\frac12) & 0 & \psi(-\frac32) & \cdots \\
0 & \psi(\frac12) & 0 & \psi(-\frac12) & 0 & \cdots \\
\psi(\frac32) & 0 & \psi(\frac12) & 0 & \psi(-\frac12) & \cdots \\
0 & \psi(\frac32) & 0 & \psi(\frac12) & 0 & \cdots \\
\vdots & \vdots & \vdots & \vdots & \vdots & \ddots
\end{pmatrix} 
\]
and the zeros sit on the integer lattice, where the bosons live.  On the doubled Hilbert space, we define
\[
C^n_{ij}=\delta_{i, j+2n}\,,
\]
where the indices \(i, j\) run over the duplicated lattice and \(n\) can be either integer or half-integer. In particular,  the operator 
\[
C^{1/2}_{ij}= \delta_{i, j+1}
\]
acts as a shift between the two lattices. The appropriate modification of the \(Q\) operator is
\[
Q_{ij}= \frac{j}{2} \delta_{ij}\,,
\]
and the translation operator on the torus becomes 
\[
U= \mathbb{I}_N \otimes \left(e^{-2\pi i Q}C^1\right)_{ij}= \mathbb{I}_N \otimes (-1)^iC^1_{ij} \, .
\]
As a consequence, anti-periodic fermions are decomposed as 
\[
\psi= \sum_{r \in \mathbb{Z}+\frac{1}{2}}\psi(r) \otimes C^r\,,
\]
and the commutation relation (\ref{eq:commutation}) (which is also valid for half-integer \(n\)) implies 
\[
(U)^{-1} \psi U= -\psi\,,
\]
as required.

This construction generalises naturally to the case of mixed boundary conditions. The fermions are decomposed into a periodic and an anti-periodic component,  the details of the decomposition depending on the specific choice of boundary conditions (we have considered different examples in the main text). Schematically 
\[
\psi = \psi^\uparrow+\psi^\downarrow\,,
\]
and our construction implies that the periodic component \(\psi ^\uparrow\) is decomposed using integer modes, thus living on the periodic lattice, while \(\psi^\downarrow\) is anti-periodic and lives on the shifted lattice. 

With these decompositions, the interaction of an uncompactified boson with a fermion takes the form
\begin{equation}
\mathrm{Tr} \,\bar\psi\Gamma^I[A_I,\psi]=\sum_{ n\in \mathbb{Z};r,p \in \mathbb{Z}+\frac{1}{2}\vert n+r+p=0} \mathrm{Tr}_Y\, \psi(r)[A(n),\psi(p)]
\end{equation}
where the trace on the right-hand side is taken over the $N$-dimensional subspace \eqref{eq:TensorHilbert}.
We see that this term has a single $U(N)$ symmetry, just as the theory with periodic fermions has. 

\section{Loop propagator regularisation for BFSS}
\label{sec:regularise}
In this section, we extend the regularisation method of \cite{Laliberte_2024} for loop propagators to the case of coefficients arising from loops in the Wilsonian effective theory analysis of compact BFSS theory.

We begin by considering the bosonic sum 
\begin{equation*}
    S_{B_1}=\sum_{n^a \in \mathbb{Z}^3} \nolimits' \int_{-\infty}^{+\infty} dp \frac{1}{p^2+n^a n_a} \, ,
\end{equation*}
and we introduce a regulator $\alpha$ such that
\begin{equation*}
    S_{B_1}=\lim_{\alpha \rightarrow 0}\sum_{n^a \in \mathbb{Z}^3} \nolimits' \int_{-\infty}^{+\infty} dp \frac{e^{-\alpha^2 n^a n_a-\alpha^2 p^2}}{p^2+n^a n_a} \, .
\end{equation*}
We rewrite this as 
\begin{equation*}
    S_{B_1}=\lim_{\alpha \rightarrow 0} \sum_{n^a \in \mathbb{Z}^3}\nolimits' \int_{-\infty}^{+\infty} dp \int_0^{+\infty} dt\,
 e^{-(t+\alpha^2)n^a n_a}e^{-(t+\alpha^2)p^2} \, ,
 \end{equation*}
 which can be rewritten in terms of the $\theta$-function
 \begin{equation*}
     \theta(t)=\sum_{n=-\infty}^{+\infty} e^{-\pi t n^2} \, ,
 \end{equation*}
 such that
 \begin{equation*}
     S_{B_1}=\lim_{\alpha \rightarrow 0} \sqrt{\pi} \int_{-\infty}^{\infty} dp \int_{\alpha^2/\pi}^{+\infty} dt e^{-tp^2}(\theta^3(t)-1) \, .
 \end{equation*}
 Focusing on the divergent part of the integral, we have 
 \begin{equation*}
     \int_{\alpha^2/\pi}^1 dt \,e^{-tp^2} (\theta^3(t)-1)=\int_{\alpha^2/\pi}^1 dt \frac{e^{-tp^2}}{t^{3/2}}(\theta^3(1/t)-1)+ F(\alpha; p) \, ,
 \end{equation*}
 where 
 \begin{equation*}
     F(\alpha;p)=\int_{\alpha^2/\pi}^1 dt \, e^{-tp^2}\left(\frac{1}{t^{3/2}}-1\right)
 \end{equation*}
 represents the divergent component of the integral. Here we made use of the property of the $\theta$-function
 \begin{equation*}
     \theta(t)=\frac{1}{t^{1/2}} \theta(1/t) \, .
 \end{equation*}
 By rearranging the various terms, we obtain
 \begin{equation*}
     S_{B_1}=\lim_{\alpha \rightarrow 0} \sqrt{\pi} \int_{-\infty}^{+\infty}dp \int_1^{+\infty} dt \left[\frac{e^{-p^2/t}}{\sqrt{t}}+e^{-p^2t}\right](\theta^3(t)-1)+\sqrt{\pi} \int_{-\infty}^{+\infty} dp F(\alpha; p) \, ,
 \end{equation*}
 where the first integral is convergent, and we regularised the divergence to be in the second term.

 We now consider the fermionic (in the case of anti-periodic boundary conditions) sum
 \begin{equation*}
     S_{F_1}=\sum_{r^a \in \mathbb{L}_3} \int_{-\infty}^{+\infty} dp \frac{1}{p^2+r^a r_a} \, ,
 \end{equation*}
 which we regulate as 
 \begin{equation*}
     S_{F_1}=\sum_{r^a \in \mathbb{Z}^3} \int_{-\infty}^{+\infty} dp \frac{e^{-\alpha^2((r^a+u^a)^2+p^2)}}{p^2+(r^a+u^a)^2}, \quad u^a=\frac{1}{2}(1,...,1) \, .
 \end{equation*}
 We can rewrite this as 
 \begin{equation*}
    S_{F_1}= \int_{-\infty}^{+\infty} dp \int_{0}^{+\infty} dt \sum_{r^a\in \mathbb{Z}^3} e^{-(\alpha^2+t)(p^2+(r^a+u^a)^2)}= \sqrt{\pi} \int_{-\infty}^{+\infty} dp \int_{\alpha^2/\pi}^{+\infty} dt \, \theta^3(t|u) e^{-tp^2} \, ,
 \end{equation*}
 where the generalized $\theta$-function takes the form
 \begin{equation*}
     \theta(t|u)=\sum_{n =-\infty}^{+\infty} e^{-\pi t(n+u)^2} \, .
 \end{equation*}
 Again, we look at the divergent part, which reads
 \begin{equation*}
     \int_{\alpha^2/\pi}^1 dt\, e^{-tp^2} \theta^3(t|\frac{1}{2})=\int_{\alpha^2/\pi}^1 dt\, \frac{e^{-tp^2}}{t^{3/2}}(e^{-\frac{3}{4}\pi t }\theta^3(\frac{1}{t}|\frac{it}{2})-1)+\tilde{F}(\alpha;p) \, ,
 \end{equation*}
 where
 \begin{equation*}
     \Tilde{F}(\alpha;p)=\int_{\alpha^2/\pi}^1 dt\, \frac{e^{-tp^2}}{t^{3/2}} 
 \end{equation*}
 and we used the property
 \begin{equation*}
     \theta(t|a)=\frac{e^{-\frac{\pi}{4} t}}{t^{1/2}}\theta(\frac{1}{t}|\frac{it}{2}) \, .
 \end{equation*}
 It follows that
 \begin{equation*}
     S_{F_1}=\sqrt{\pi} \int_{-\infty}^{+\infty} dp \int_{1}^{+\infty}dt \left(\frac{e^{-p^2/t}}{\sqrt{t}}(e^{-\frac{3}{4}\pi /t}\theta^3(t|\frac{i}{2t})-1)+e^{-p^2t}\theta^3(t|\frac{1}{2})\right)+\sqrt{\pi} \int_{-\infty}^{+\infty} dp \,\tilde{F}(\alpha;p) \, ,
 \end{equation*}
 where the second term is the divergent one.

As a consequence, when we consider the difference $(S_{F_1}-S_{B_1})$, the divergent parts cancel out, giving
\begin{equation*}
    \int_{-\infty}^{+\infty} dp \left(\tilde{F}(\alpha;p)-F(\alpha;p) \right)=\int_{\alpha^2/\pi}^1dt\int_{-\infty}^{+\infty} dp\,e^{-tp^2}=2 \sqrt{\pi} \, ,
\end{equation*}
such that the difference reads
\begin{multline*}
    S_{F_1}-S_{B_1}=2\pi-\sqrt{\pi}\int_{-\infty}^{+\infty} dp\int_{1}^{+\infty}dt\left[\left(\frac{e^{-p^2/t}}{\sqrt{t}}+e^{-p^2t}\right)(\theta^3(t)-1)\right. \\ \left.
    -\left(\frac{e^{-p^2/t}}{\sqrt{t}}(e^{-\frac{3\pi}{4t}}\theta^3(t|\frac{i}{2t})-1)+e^{-p^2t}\theta^3(t|\frac{1}{2})\right)\right] \,,
\end{multline*}
which is a finite number and can be evaluated numerically.
\section{Filling up the Fermi sea}
\label{sec:fermisea}
In this section, we focus on the fermionic Hamiltonian (\ref{eq:Fermenergy}), which represents a collection of fermionic oscillators. We show how the Fermi sea should be filled up in order to interpret the fuzzy sphere solution as a black hole.

We begin by noting that $|\lambda_{-,n-1}|=\lambda_{+,n}$, so that for $\lambda_{+,1}=1\leq E_n\leq|\lambda_{-,N-2}|=N-1$ the shell with energy $E_n$ (ignoring the prefactor) is composed of the $4(2n+2)$ oscillators $\xi^{j, n}_\alpha$ and the $4(2n-2)$ oscillators $\chi^{j, n-1}_\alpha$, giving total dimension $\#(E_n)=16n$. It follows that the shell with the largest dimension is the one with energy $E_{N-1}=|\lambda_{+, N-1}|$, having dimension $\#(E_{N-1})=4(4N-4)$. Moreover, the state with the largest degeneracy is the one for which such a shell is half-filled, which yields
\[
\Omega=\binom{16N-16}{8N-8}\propto \frac{2^{16(N-1)}}{\sqrt{N}}\, .
\]
At leading order in the large $N$ expansion, we can compute the fermionic energy contribution from the shell with energy $E_{N-1}$. This follows straightforwardly from (\ref{eq:Fermenergy}). In particular, if we are free to excite $j=4(1+x)(2N-2)$ oscillators, $-1\leq x\leq1$ being the fraction of excited oscillators. The energy reads 
\[
H_{ferm}=\frac{8R\,k^4}{3}N^3+O(N^2)=\frac{32R \, k^2}{3}N \, R^2_{f}+O(N^2)
\]
where $R_f=\frac{k\,N}{2}$ is the physical fuzzy sphere radius in the large $N$ limit. At the same time, the bosonic part of the Hamiltonian yields, when evaluated on the fuzzy sphere background 
\[
H_{bos}=-\frac{R\, k^2}{2}NR^2_f
\]
So the total energy reads 
\[
H=\frac{R\, k^2}{6}(64-3)NR^2_f \, .
\]
As explained in section \ref{sec:BHsol}, the Hamiltonian gives the lightcone energy, and it is related to the invariant mass through 
\[
H=E_{LC}=\frac{R}{2N}M^2 \,.
\]
The mass-radius relation  
\[
M_{5D}=r_0^2
\]
then implies 
\[
r_0=\left(\,\frac{244}{3}\right)^{\frac{1}{4}}R_f \, .
\]
In order for the fuzzy sphere geometry to represent a black hole, its radius must be smaller than the Schwarzschild radius. We see that this condition is satisfied when the Fermi sea is filled in this way. The fraction \(x\) of excited oscillators in the \(E_N\) determines the entropy.
In the main text, we will express the results in terms of the fraction \(y\) defined as 
\[
y=\frac{1+x}{2} , \qquad 0 \leq y \leq 1.
\]

\section{Stability of the squashing modes}
\label{sec:stability}
In this section, we prove that the squashing modes of bosonic fluctuations are stabilised by the fermionic sector. We have seen in section \ref{sec:bosonicfluc} that quadratic bosonic fluctuations with \(\ell=1, J=2\) are unstable. These correspond to the squashing modes of the fuzzy sphere 
\[
\delta X^i = S^i_j X^j \, ,
\]
where \(S^i_j\) is a symmetric traceless tensor and \(X^j\) are the bosonic matrices for the fuzzy sphere background (\(SU(2)\) generators). The quadratic potential reads 
\[
\Delta V_{bos}= R k^4\Tr \left[\frac{1}{2} \delta X^i N_{ij} \delta X^j\right]
\]
and, given that those are eigenmodes of \(N_{ij}\) with eigenvalue \(-2\) we get
\[
\Delta V_{bos}=-R k^4 \, S^i_j S^i_k \Tr \left[X^j  X^k\right] \, .
\]
Moreover, using 
\[
\Tr \left[X^j  X^k\right] =\frac{N(N^2-1)}{12} \delta^{jk}
\]
we obtain 
\[
\Delta V_{bos}= -R k^4 \frac{N^3}{12} \Tr[S^2]
\]
at leading order in \(N\). The negative sign corresponds to an instability. We will now see how this is compensated by including quadratic perturbations, due to the same modes, from the fermionic sector. The perturbation to the fermionic operator, due to the squashing modes, reads 
\[
\delta O_f=Rk^4 \, \Tr\left[\lambda^{\uparrow, \dagger} \sigma^i [\delta X^i, \lambda^\uparrow] \right]=Rk^4 \,\Tr\left[\lambda^{\uparrow, \dagger} \sigma^i S^i_j L^j\lambda^{\uparrow}\right], \quad L^j=[X^j, \cdot] \, .
\]
We will compute the effective potential generated by the fluctuations using second-order perturbation theory on the state of the system, which we recall is the Fermi sea with partially filled highest shell. To this end, it is useful to decompose the fermions as 
\begin{equation}
    \label{eq:fermion_creation}
\lambda^\uparrow_{b \beta \sigma, \alpha}=U^{r,n, \dagger}_{b\beta \sigma} \xi^{r,n}_\alpha+ V^{r,n,\dagger}_{b\beta \sigma}\chi^{r,n, \alpha \dagger} \, ,
\end{equation}
such that 
\[
\delta O_f = R k^4 \left(P^{r, q}_{nm}\xi^{r,n, \alpha\dagger}\xi^{q,m}_\alpha-Q^{r, q}_{nm}\chi^{r,n, \alpha \dagger}\chi^{q,m}_\alpha+T^{r, q}_{nm}\xi^{r,n, \alpha\dagger}\chi^{q,m, \alpha \dagger}+T^{r, q \,\dagger}_{nm}\chi^{q,m }_\alpha\xi^{r,n}_\alpha\right)
\]
with 
\[
P^{r, q}_{nm} := U^{r,n } A U^{q,m, \dagger}, \quad Q^{r, q}_{nm} := V^{r,n } A V^{q,m,\dagger}, \quad T^{r, q}_{nm} := U^{r,n} A V^{q,m, \dagger}, \quad A:=\sigma^i S^{i}_j L^j \, ,
\]
and, in the last expression, we suppressed matrix indices for simplicity of notation. 

We will compute the Hamiltonian at second order in perturbation theory. First, we will show that the first-order term vanishes. This reads 
\[
\Delta V_{fo}=\frac{1}{\Omega} \sum_{S \in \mathcal{H}_0} \langle S | \delta O_f | S \rangle \, ,
\]
where \(\mathcal{H}_0\) is the Hilbert space of the state of the system, composed of the Fermi sea with partially filled highest shell, and \(\Omega\) is the degeneracy. Hence, \(|S \rangle\) are states in this Hilbert space. The only terms which will contribute are those that preserve the particle number. Hence, we get non-trivial contributions from 
\[
\langle S |\xi^{r,n, \alpha\dagger}\xi^{q,m}_\alpha | S \rangle= 4 \delta^{rq} \delta^{nm} \, ,
\]
and similarly for \(\chi\) oscillators. It follows that 
\[
\Delta V_{fo}=4 R k^4\sum_n \sum_r\left(P^{r,r}_{n,n}-Q^{r,r}_{n,n}\right)
\]
which can be rewritten as 
\[
\Delta V_{fo}=4 R k^4\sum_n \Tr\left[P_{+,n}A-P_{-,n}A\right] \, ,
\]
where 
\[
P_{+,n}= \frac{\mathrm{X}_n+(n+1)}{2n+1}, \qquad P_{-,n}= \frac{n-\mathrm{X}_n}{2n+1}
\]
are the projectors in the \(+\) and \(-\) sectors, constructed from the operator \(\mathrm{X}_n=(\sigma^i L_i)|_n\). We can use 
\[
\begin{aligned}
    \Tr\left(A\right)&=0 \\
    \Tr\left(\sigma^i \sigma^j\right)&=2 \delta^{ij} \\
    \Tr\left(L^i L^j\right)&=\frac{1}{3}n(n+1)(2n+1) \delta^{ij}
\end{aligned}
\]
to show that
\[
\Delta V_{fo} \propto \Tr \left(S\right)=0 \, .
\]

We now proceed to compute the effective Hamiltonian generated at quadratic order by the perturbation. This is determined by 
\[
\Delta V_{ferm}= \frac{1}{\Omega } \sum_{S \in \mathcal{H}_0} \sum_{R \neq S}\langle S | \delta O_f | R \rangle \frac{1}{E_S-E_R} \langle R | \delta O_f | S \rangle \, ,
\]
 and we are summing over all possible states that can be connected by the action of \(\delta O_f\). The denominator gives the energy difference between the two states. Let us examine the action of \(\delta O_f\) on states \(| S \rangle\). The leading contribution in the large \(N\) expansion is obtained by acting with creation and annihilation operators on all the levels of the Fermi sea. When the action only involves the top, half-filled level, the corresponding contribution will be subleading in \(N\). The operator \(\xi^{r,n, \alpha\dagger}\xi^{q,m}_\alpha\) removes and adds one oscillators. This requires \(r=q\) and \(n=m\). As a consequence, this would impose \(R=S\), which does not appear in the sum. Hence, this term never contributes, at leading order in \(N\). The same applies to \(\chi^{r,n, \alpha\dagger}\chi^{q,m}_\alpha\). The action of \(\xi^{r,n, \alpha\dagger}\chi^{q,m, \alpha \dagger}\) creates oscillators on states which are already filled, and therefore provides a vanishing contribution. The only non-zero term comes from removing two fermions from the sea by the action of \(\chi^{q,n}_\alpha\xi^{r,m}_\alpha\).  
Moreover, \(P, Q, T\) terms are constructed from the interpolation of the operator \(A\) between different eigenstates. But \(A\) can only interpolate between states with the same orbital angular momentum \(n\). This imposes \(n=m\) on all \(P, Q, T\) factors. It follows that the only non-zero term is \(T^{r,q \,\dagger}_{n, n} \chi^{q,n}_\alpha\xi^{r,n}_\alpha\). This selects a state \(|R \rangle\) where two oscillators have been removed. The corresponding energy difference is 
\[
E_S - E_R=R k^4( \lambda_{+, n}- \lambda_{-,n})=R k^4(2n+1) \, .
\]
Putting all together, we get 
\[
\Delta V_{ferm}=4R k^4\sum_n \frac{|T_{n,n}^{r,q}|^2}{2n+1}\, ,
\]
where the factor of \(4\) from summing over \(\alpha=1,\dots4\). Hence, computing the fermionic correction reduces to the evaluation of the factor 
\[
|T_{n,n}^{r,q}|^2=\Tr\left[P_{+,n} A P_{-,n}A\right] \, .
\] 
Expanding the expression, we get
\[
|T_{n,n}^{r,q}|^2=\frac{1}{(2n+1)^2}[n(n+1) \Tr\left(A^2\right)-\Tr \left(\mathrm{X}A^2\right)-\Tr \left(\mathrm{X}A\mathrm{X}A\right)] \, .
\]
The traces can be evaluated using standard \(SU(2)\) identities and read 
\[
\begin{aligned}
    \Tr\left(A^2\right)&=\frac{2}{3}n(n+1)(2n+1) \Tr \left[S^2\right] \\
     \Tr\left(\mathrm{X}A^2\right)&=\frac{1}{3}n(n+1)(2n+1) \Tr \left[S^2\right] \\
      \Tr\left(\mathrm{X}A\mathrm{X}A\right)&=-\frac{2}{15}n(n+1)(2n+1)(n-1)(n+2) \Tr \left[S^2\right] \, .
\end{aligned}
\]
Putting everything together 
\[
|T_{n,n}^{r,q}|^2=\frac{n(n+1)(2n-1)(2n+3)}{5(2n+1)} \Tr \left[S^2\right] \, .
\]
It follows that 
\[
\sum_n \frac{|T_{n,n}^{r,q}|^2}{(2n+1)}=\frac{N^3}{15} \Tr \left[S^2\right] + O(N^2) \, ,
\]
and, as a consequence
\[
\Delta V_{ferm}= Rk^4 \frac{4N^3}{15}\Tr \left[S^2\right] +O(N^2) \, .
\]
Considering both bosonic and fermionic contributions, the quadratic perturbation potential for the squashing modes reads 
\[
\Delta V_{squash}= \Delta V_{bos} + \Delta V_{ferm}= Rk^4 N^3 \left(-\frac{1}{12}+\frac{4}{15}\right)\Tr \left[S^2\right] +O(N^2)=Rk^4 N^3 \frac{11}{60}\Tr \left[S^2\right] +O(N^2)
\]
which is positive. Hence, the perturbation is stable.

A similar analysis can be carried out for the \(\ell=1\) modes in the directions orthogonal to the fuzzy sphere. These modes can be written in terms of fuzzy sphere matrices as
\[
\delta X^{i+3}= A^i_j X^j \, ,
\]
where \(A^i_j\) is the \(3 \times 3\), anti-symmetric matrix parametrising the fluctuation. The analogous analysis to the previous case can be carried out using the fermionic term 
\[
\delta O_f ^\prime = i\lambda^{\uparrow, T}(t,0)i\sigma^2 \gamma^{i+3} A^i_j[X_j, \lambda^\uparrow(t,0)]
\]
obtained by perturbing the corresponding term in the full action (\ref{eq:BFSS_Mixed_eff}). This fixes the instability of the \(\ell=1\) mode in the orthogonal directions.
\bibliographystyle{JHEP}
\bibliography{references.bib}
\end{document}